\documentclass[sigplan,10pt]{acmart}
\renewcommand\footnotetextcopyrightpermission[1]{}

\usepackage{pbalance}   

\usepackage{tikz}
\usepackage{amsmath}
\usepackage{orcidlink}

\usepackage{xcolor}
\usepackage{xifthen}
\usepackage[normalem]{ulem}
\usepackage[capitalize]{cleveref}
\usepackage{mathrsfs} \usepackage{listings}

\usepackage{enumitem}
\setdescription{leftmargin=15pt,itemsep=0pt}

\definecolor{color0}{RGB}{0, 0, 0} \definecolor{color1}{RGB}{255, 65, 68} \definecolor{color1b}{RGB}{252, 91, 99} \definecolor{color2}{RGB}{234, 127, 43} \definecolor{color2b}{RGB}{253, 174, 97} \definecolor{color2c}{RGB}{203, 124, 97} \definecolor{color3}{RGB}{200, 200, 0} \definecolor{color3b}{RGB}{255, 255, 191} \definecolor{color4}{RGB}{98, 181, 86} \definecolor{color4b}{RGB}{171, 221, 164} \definecolor{color4c}{rgb}{0.31, 0.47, 0.26}
\definecolor{color5}{RGB}{43, 131, 186} \definecolor{color5b}{RGB}{158, 204, 239} 

\newcommand{\gennote}[4][blue]{\textcolor{#1}{$\rule{8pt}{8pt}_\textsf{\scshape\bfseries #2}$ \textcolor{gray}{\emph{\sout{#3}}}#4}}

\newcommand{\authorcomment}[3]{\expandafter\newcommand\csname#1\endcsname[2][]{\gennote[#3]{#2}{##1}{##2}}}
\renewcommand{\authorcomment}[3]{\expandafter\newcommand\csname#1\endcsname[2][]{}}

\newcommand{\id}[1]{\textit{#1}}
\newcommand{\const}[1]{\textsf{#1}}
\newcommand{\eventname}[1]{\const{#1}}

\newcommand{\condparenth}[1]{\ifthenelse{\isempty{#1}}{}{(#1)}}
\newcommand{\condspace}[1]{\ifthenelse{\isempty{#1}}{}{\,#1}}
\newcommand{\condsetparenth}[1]{\ifthenelse{\isempty{#1}}{}{\{#1\}}}
\newcommand{\condsub}[1]{\ifthenelse{\isempty{#1}}{}{_{#1}}}
\newcommand{\condcdot}[1]{\ifthenelse{\isempty{#1}}{~\cdot~}{#1}}

\newcommand{\projection}[2]{\textit{#1}\condparenth{#2}}

\newcommand{\external}[1]{\projection{ext}{#1}}
\newcommand{\internal}[1]{\projection{int}{#1}}
\newcommand{\prev}{\id{prev}}
\newcommand{\nxt}{\id{next}}

\newcommand{\authorizedcore}[1]{\textit{auth-core}}
\newcommand{\authorizeddown}[1]{\textit{auth-down}}
\newcommand{\authorizedup}[1]{\textit{auth-up}}
\newcommand{\authorized}[1]{\textit{auth}}

\newcommand{\CustProv}{\textit{CustProv}}
\newcommand{\ProvCust}{\textit{ProvCust}}
\newcommand{\Core}{\textit{Core}}

\newcommand{\segid}[1]{\ensuremath{\id{segID}_{#1}}}
\newcommand{\dir}{\ensuremath{\id{dir}}}

\newcommand{\pkt}{\id{pkt}}

\newcommand{\dispatchevent}{\eventname{Attack}}
\newcommand{\forwardevent}{\eventname{Forward}}
\newcommand{\recvevent}{\eventname{Recv}}

\newcommand{\sendevent}{\eventname{Send}}

\newcommand{\incheck}{\id{recv}}
\newcommand{\ifsvalid}{\id{ifs\_valid}}
\newcommand{\cryptovalid}{\id{crypto\_valid}}

\newcommand{\upd}{\id{upd}}
\newcommand{\add}{\id{send}}

\newcommand{\forwardmethod}{\code|process|}
\newcommand{\forwardmethodgobra}{\texttt{process}} 
\newcommand{\processmethod}{\code|process|}

\newcommand{\asinterface}{AS interface}

\newcommand{\asinterfacevar}{\code|i|}
\newcommand{\asinterfacevargobra}{\texttt{i}}

\newcommand{\inasinterfacevargobra}{\texttt{i}}
\newcommand{\outasinterfacevarname}{j}

\newcommand{\outasinterfacevargobra}{\texttt{j}}

\newcommand{\packetvar}{\code|pkt|}
\newcommand{\packetvargobra}{\texttt{pkt}}

\newcommand{\packetabstraction}{\code|Abs|}
\newcommand{\packetabstractiongobra}{\texttt{Abs}}

\newcommand{\AuthSeg}{\id{AuthSeg}}

\newcommand{\DY}{\mathsf{DY}}
\newcommand{\xor}{\oplus}

\newcommand{\eventhead}[1]{#1\!:\,}
\newcommand{\eventsep}{\triangleright}
\newcommand{\eventsepbefore}{}
\newcommand{\eventsepafter}{\hspace*{0.5mm}\triangleright\hspace{1mm}}
\newcommand{\event}[2]{\eventhead{#1} #2}

\lstdefinelanguage{gobra}{
  language=go,
  sensitive=true,
  morecomment=[l]{//},
  morecomment=[s]{/*}{*/},
  morekeywords=[1]{ pred, implements, ghost, set
  },
  morekeywords=[2]{ requires, ensures, invariant, req, ens, pure,  unfolding, in, forall, acc, decreases, old
  },
  morekeywords=[3]{ fold, unfold,
    assume, assert, inhale, exhale
  },
  basicstyle={\ttfamily\footnotesize},
  commentstyle={\color[HTML]{747678}\textit},
  morecomment=*[l][keyworsstyle]{//@},
  keywordstyle={[1]\color[HTML]{0005FF}},keywordstyle={[2]\color[HTML]{CC5500}},keywordstyle={[3]\color[HTML]{EC008C}},mathescape=true,
  columns=fullflexible,
moredelim=**[is][\normalfont\itshape]{'}{'}
}

\lstnewenvironment{gobra}[1][]{
  \lstset{
    language=gobra,
    floatplacement={tbp},
    captionpos=b,
    frame=lines,
    numbers=left, numberblanklines=false,
    xleftmargin=8pt,
    xrightmargin=8pt,
    numberstyle=\scriptsize,
    breaklines=true,
    #1
  }
}{}

\def\code{\lstinline[language=gobra,basicstyle=\ttfamily]}

\begin{document}

\date{}

\title{Protocols to Code: Formal Verification of a Next-Generation Internet Router}

\newcommand{\ethz}{Department of Computer Science, ETH Zurich}

\orcid{0000-0002-6505-3942}

\author{Jo{\~{a}}o C. Pereira}
\affiliation{
   \department{Department of Computer Science}
   \institution{ETH Zurich}
   \country{Switzerland}}
\orcid{0000-0003-4671-4132}
\authornote{Corresponding author. Email: \texttt{joao.pereira@inf.ethz.ch}}

\author{Tobias Klenze}
\affiliation{
   \department{Department of Computer Science}
   \institution{ETH Zurich}
   \country{Switzerland}}
\orcid{0000-0002-6505-3942}

\author{Sofia Giampietro}
\affiliation{
   \department{Department of Computer Science}
   \institution{ETH Zurich}
   \country{Switzerland}}

\author{Markus Limbeck}
\affiliation{
   \department{Department of Computer Science}
   \institution{ETH Zurich}
   \country{Switzerland}}

\author{Dionysios Spiliopoulos}
\affiliation{
   \department{Department of Computer Science}
   \institution{ETH Zurich}
   \country{Switzerland}}

\author{Felix A. Wolf}
\affiliation{
   \department{Department of Computer Science}
   \institution{ETH Zurich}
   \country{Switzerland}}
\orcid{0000-0002-8573-2387}

\author{Marco Eilers}
\affiliation{
   \department{Department of Computer Science}
   \institution{ETH Zurich}
   \country{Switzerland}}
\orcid{0000-0003-4891-6950}

\author{Christoph Sprenger}
\affiliation{
   \department{Department of Computer Science}
   \institution{ETH Zurich}
   \country{Switzerland}}
\orcid{0000-0003-2941-5165}

\author{David Basin}
\affiliation{
   \department{Department of Computer Science}
   \institution{ETH Zurich}
   \country{Switzerland}}
\orcid{0000-0003-2952-939X}

\author{Peter M{\"{u}}ller}
\affiliation{
   \department{Department of Computer Science}
   \institution{ETH Zurich}
   \country{Switzerland}}
\orcid{0000-0001-7001-2566}

\author{Adrian Perrig}
\affiliation{
   \department{Department of Computer Science}
   \institution{ETH Zurich}
   \country{Switzerland}}
\orcid{0000-0002-5280-5412}

\begin{abstract}
We present the first formally-verified Internet router, which is part of the SCION Internet architecture. SCION routers run a cryptographic protocol for secure packet forwarding in an adversarial environment. We verify both the protocol's network-wide security properties and low-level properties of its implementation. More precisely, we develop a series of protocol models by refinement in Isabelle/HOL and we use an automated program verifier to prove that the router's Go code satisfies memory safety, crash freedom, freedom from data races, and adheres to the protocol model.
Both verification efforts are soundly linked together.

Our work demonstrates the feasibility of coherently verifying a critical network component from high-level protocol models down to performance-optimized production code, developed by an independent team. In the process, we uncovered critical bugs in both the protocol and its implementation, which were confirmed by the code developers, and we strengthened the protocol's security properties. This paper explains our approach, summarizes the main results, and distills lessons for the design and implementation of verifiable systems, for the handling of continuous changes, and for the verification techniques and tools employed.

\end{abstract}

\settopmatter{printacmref=false, printccs=false, printfolios=true}

\maketitle
\pagestyle{plain}

\vspace{-.8ex}

\section{Introduction}

Faulty software poses a serious threat to the reliability and security of critical computing infrastructures and  may lead to catastrophic system failures and devastating attacks. 
In light of this problem, substantial progress has been made over the past decades on the theory and tools for software verification and several major software verification projects have been completed in different areas. 
In the area of operating systems, the seL4 project~\cite{kleinehacdeeknstw09} pioneered such large verification efforts, followed later by CertiKOS~\cite{DBLP:conf/osdi/GuSCWKSC16}. 
There are also verified compilers, including CompCert~\cite{DBLP:journals/cacm/Leroy09} and CakeML~\cite{DBLP:conf/popl/KumarMNO14}. 
In the area of distributed systems, IronFleet~\cite{dblp:conf/sosp/hawblitzelhklpr15}, Verdi~\cite{dblp:conf/pldi/wilcoxwptwea15}, and Velisarios~\cite{dblp:conf/esop/rahlivvv18}, respectively verified the Paxos, Raft, and PBFT distributed consensus protocols.
The Everest project successfully verified security protocols~\cite{DBLP:conf/sp/BhargavanFKPS13,DBLP:conf/sp/Delignat-Lavaud17,DBLP:conf/sp/Delignat-Lavaud21} and cryptographic libraries~\cite{DBLP:conf/ccs/ZinzindohoueBPB17,DBLP:conf/sp/ProtzenkoPFHPBB20} at the code level.

Given that Internet routers are a central part of our critical networking infrastructure and deployed on a large scale, their security and reliability are of paramount importance. In this paper, we present the first comprehensive formal verification of Internet routers and their protocols. 

Today's Internet routers run the aging Border Gateway Protocol (BGP), 
which does not meet modern security and reliability requirements.
Therefore, our verification effort focuses on the SCION Internet architecture~\cite{scionbookv2}, a clean-slate redesign of an inter-networking infrastructure with a focus on strong security and reliability properties. SCION border routers are responsible for packet forwarding, which is simple and efficient, since each packet's path is embedded in the packet header, alongside cryptographic authenticators that authorize the path.  
SCION has real-world traction. It is deployed and offered by 14
major Internet service providers reaching networks on 5 continents and is used, for instance, as the networking layer for Swiss interbank clearing \cite{SSFN-SIX}.

The security and correctness of Internet routing depend on both \emph{system-wide, global, security properties of the protocol} (e.g., that packets can travel only along previously authorized paths, a property called \emph{path authorization}) and \emph{local properties of the code} (e.g., that the protocol is implemented correctly). 
Our work provides strong guarantees for both protocol and code. SCION's simplicity is a key enabling factor for its verification. 
Nonetheless, the verification itself raises several challenges.

On the protocol side, we must account for the complex adversarial networking environment in which SCION routers are intended to operate. This involves an arbitrary network topology, an arbitrary set of authorized paths, as well as active and possibly colluding attackers.
Moreover, the protocol is designed to achieve global properties like path authorization and loop freedom. Efficiency mandates that each router performs only local (cryptographic and other) checks related to its own position on a packet's embedded path. 
Showing that these checks imply the global properties requires stating and proving suitable invariants over recursive data structures that hold despite the attackers' presence.
In contrast, the verification of network configurations (see, e.g., \cite{DBLP:conf/aplas/Kozen14,
LiuHSSLSWCMF18,
DBLP:conf/oopsla/WeitzWTEKT16,
DBLP:conf/sigcomm/BeckettGMW17}) considers a non-adversarial setting and either single nodes or fixed, concrete network topologies.

On the implementation side, we face the challenge that the router code was developed independently of our verification effort, with an emphasis on performance.
Accordingly, the SCION router is implemented in roughly 4.700
lines of optimized, concurrent Go code that heavily uses aliasing to minimize memory consumption and copying. Verifying this implementation requires developing a verification technique and tool that supports all used language features and programming idioms.
Finally, our verification approach must be capable of soundly transferring the global protocol properties proven for the design to the code level such that they also hold for the executing system.

All the verification projects previously mentioned employ commonly used, effective techniques to facilitate the verification task, such as the co-development of the implementation and its verification, the adoption of a programming language and a software design that ease verification, and code extraction from correctness proofs. However, our setup comes with constraints and challenges that prevent the application of these techniques and require a novel verification approach.

\paragraph{Approach}

\looseness=-1
We employ the Igloo methodology~\cite{10.1145/3428220}, which soundly combines protocol verification by refinement with code verifiers based on separation logic~\cite{DBLP:conf/lics/Reynolds02}. 
To verify the SCION router, we model the protocol and its adversarial networking environment as a labeled transition system in Isabelle/HOL \cite{dblp:books/sp/nipkowpw02}. We start from an abstract protocol under a weak adversary and refine it into more concrete ones under a strong symbolic (Dolev-Yao~\cite{dblp:journals/tit/dolevy83}) adversary.  
For these models, we prove global properties, including path authorization and loop freedom. 
From the most concrete model, we automatically generate a program specification for the router, which completely describes the router's allowed I/O behavior. 
We then use the Gobra verifier~\cite{DBLP:conf/cav/WolfACOPM21} to verify the router's Go code against this specification.  This ensures that the code correctly follows the protocol and establishes additional code properties such as memory safety, crash freedom, 
and freedom from data races. Since verification is performed statically, the executable code and its performance are entirely unaffected.
Igloo's soundness result implies that the protocol properties are preserved for the code executing in its adversarial network environment.
We build on previous work~\cite{klenzecsf,KlenzeSprengerBasin-JCS-2022}, which modeled a simplified version of the SCION protocol with uni-directional single-segment paths and verified only path authorization, without verifying the code.

\paragraph{Contributions}
We summarize our main contributions. 
\begin{itemize}
\item We present the first comprehensive
verification of a full-fledged Internet router, and indeed of any networking infrastructure component. We have proved both system-wide protocol security properties and local code properties. In the process, we discovered and fixed errors in the protocol and its implementation, despite extensive prior reviewing and testing.

\item We demonstrate the feasibility of verifying advanced, performance-optimized production code developed by a different team. This required new verification techniques for the language features and software designs used by this program, as well as substantial performance improvements to Gobra to handle the complexity of production code.

\item We summarize our experience in verifying a complex system independently from its development. In particular, we explain how we use different forms of modularity to deal with continuous changes and to structure the overall task into manageable chunks.
\end{itemize}

\section{Background: SCION overview}
\label{sec:scion}

\looseness=-1
SCION is a secure inter-domain network architecture, providing connectivity between autonomous systems (AS).
The control plane performing path discovery and dissemination executes on a service infrastructure, whereas the data plane providing packet forwarding runs on border routers.
Although this paper presents the verification of the border router, this section provides an overview of the entire architecture.

\begin{figure}[t]
    \centering
    \includegraphics[width=6cm]{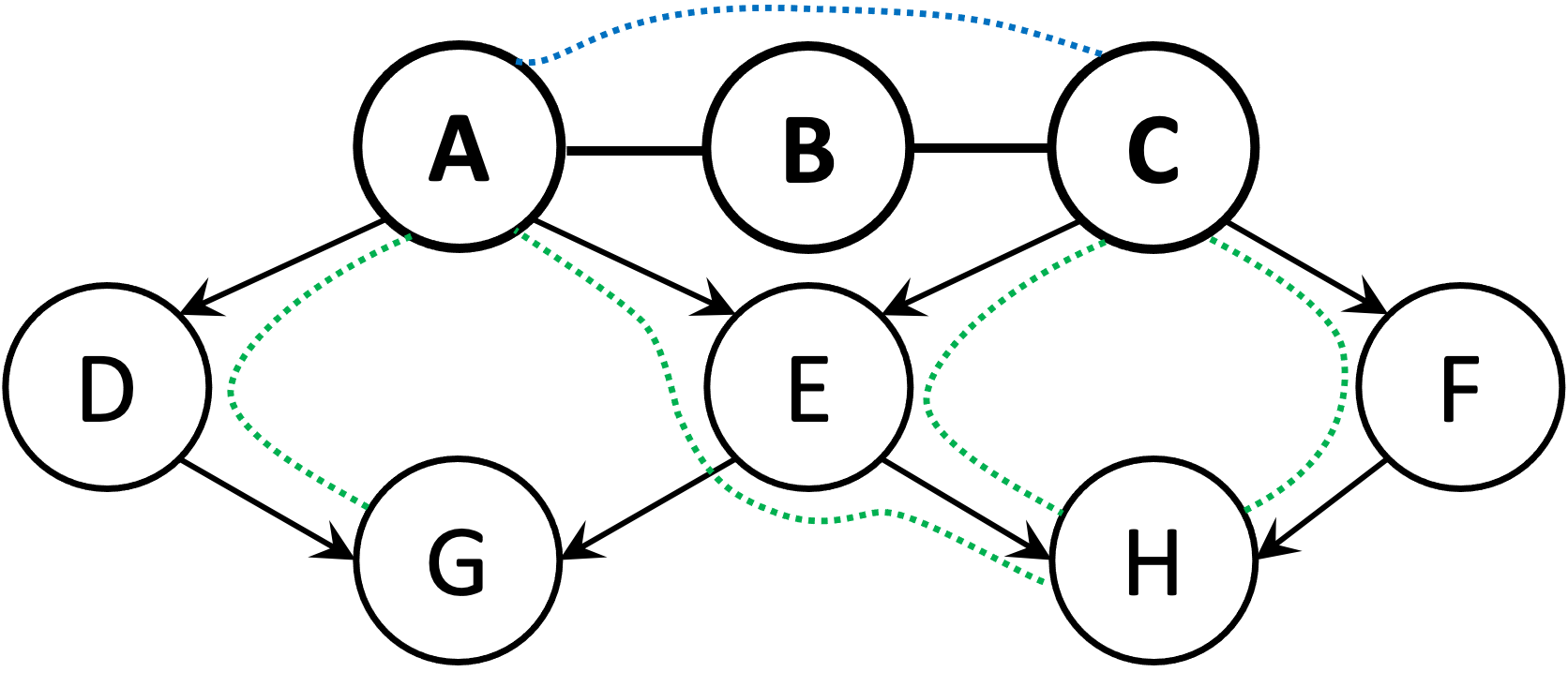}
    \caption{Network topology example. Autonomous systems are linked hierarchically (arrows), except among cores ASes (bold). Dotted lines represent authorized path segments.}
\label{fig:topology}
\end{figure}

\paragraph{SCION network} 
ASes in the SCION network are organized hierarchically, where each AS has bidirectional links to neighboring ASes: \CustProv{} links to provider ASes and \ProvCust{} links to customer ASes. 
ASes at the top of the hierarchy, i.e., without providers, are \emph{core ASes}, which connect to each other via \Core{} links. 
\Cref{fig:topology} shows an example.

Every AS assigns to each of its inter-AS links an \emph{interface}, which, in combination with the AS identifier, is globally unique. SCION routers are \emph{border routers} as they connect the internal AS network to other ASes via inter-AS links.

\paragraph{Segment construction.}
The control plane discovers and simultaneously authorizes forwarding paths between ASes. For scalability reasons, paths are established along several \emph{authorized segments}. \emph{Down-segments} are constructed along \ProvCust{} links starting at a core AS, 
and \emph{core segments} along \Core{} links. 
\emph{Up-segments} are obtained by reversing down-segments. Core segments can also be reversed for the traversal in the opposite direction.
All segments are authorized by the on-path ASes using nested message authentication codes (MACs), which we will come back to in \cref{ssec:protocol}.

\looseness=-1 Segments consist of a sequence of \emph{hop fields}, each carrying the forwarding information of one AS. Each hop field contains the interfaces of the incoming and outgoing links (\emph{prev} and \emph{next}), and a cryptographic \emph{authenticator} containing the nested MACs, which 
the respective router uses to verify that the segment was authorized by the control plane. 
Crucially, not only the adjacent inter-AS links, but the entire segment is authorized. This rules out subtle \emph{path splicing} attacks~\cite{KlenzeSprengerBasin-JCS-2022}.  
\paragraph{Segment combination. }
To send a packet, the source end host combines one or more segments to form an AS-level path from its own AS to the destination end host's AS, and embeds this path in the packet header. 
Such a path can consist of up to three segments: 
an up-segment from the source AS to a core AS, a core-segment to another core AS, 
and a down-segment to the destination AS.

The combination of segments follows rules that protect the economic interests of ASes. One central goal is to avoid \emph{valleys}, where an AS forwards packets from a provider to a provider (e.g., from $A$ to $C$ via $E$ in \cref{fig:topology}). Since ASes are paid by their customers, and must pay their providers, such packets only create costs but no revenue for the valley AS.

\paragraph{Forwarding} In the data plane, routers forward each packet along the sender-selected path, after validating the AS's cryptographic authenticator. 
Inter-domain forwarding tables are not required since each packet contains its own forwarding state.
Note that routing and forwarding within each AS are performed by the AS's intra-domain routing system and are not part of SCION\@.
We further explain the forwarding protocol in \cref{sec:protocol-verification}.

\section{Overview of verification }
\label{sec:overview}

We verify SCION's data plane and its implementation in routers using the Igloo framework~\cite{10.1145/3428220}, which allows us to verify the protocol and the code separately, while still obtaining global guarantees for the executing system. This separation of concerns enables us to leverage the strongest and most appropriate techniques and tools available for each purpose: we use stepwise refinement in Isabelle/HOL for the protocol and automated deductive verification in Gobra for the implementation. 

\begin{figure}[t]
    \centering
    \includegraphics[width=6cm]{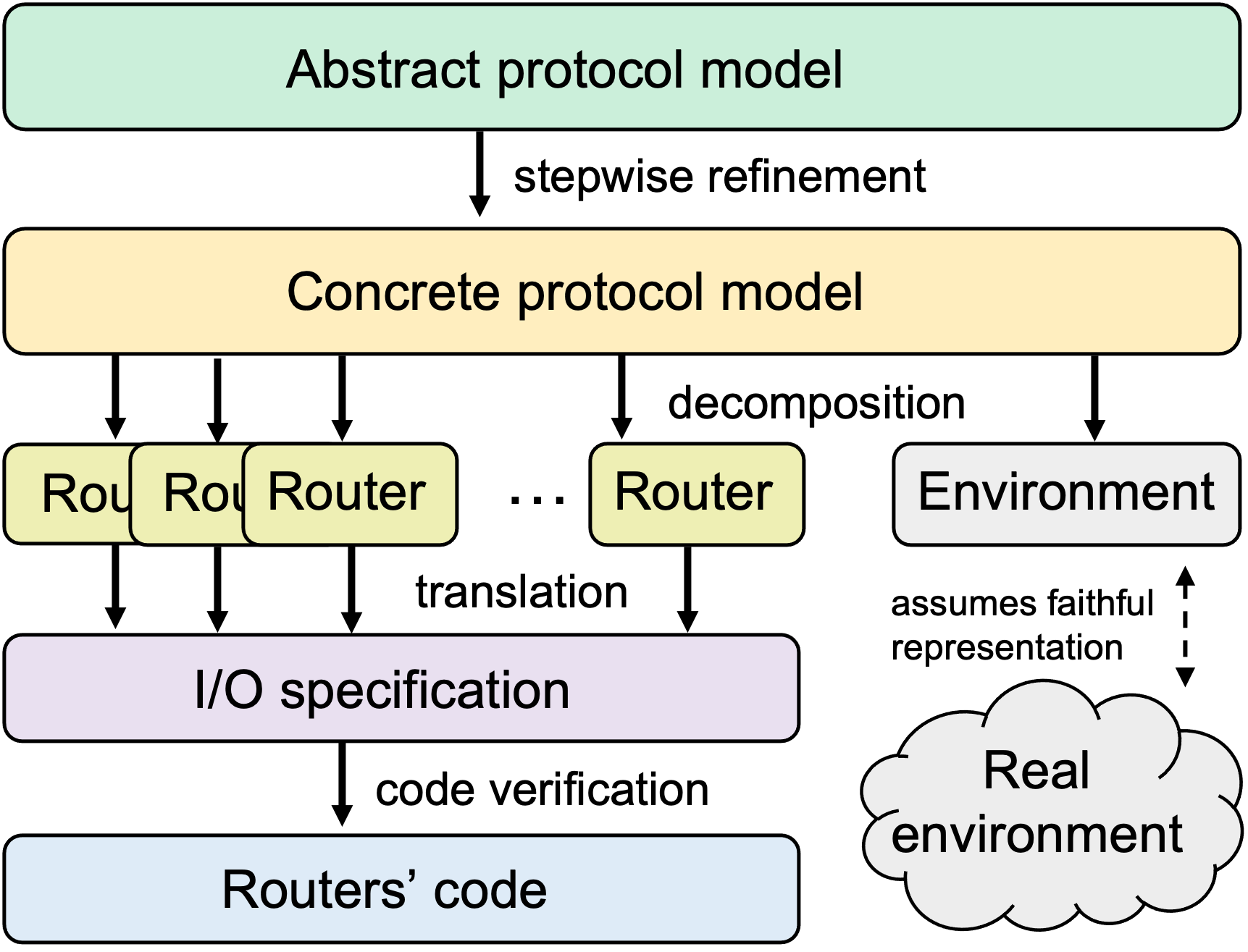}
    \caption{Overview of our approach.}
\label{fig:overview}
\end{figure}

Below, we summarize the properties we establish, give an overview of the protocol and code verification (cf.~\cref{fig:overview}), and state the assumptions on which our verification effort relies. \Cref{sec:protocol-verification,sec:code-verification} provide more details about the verification.

\subsection{Verification guarantees}

On the protocol level, we prove that SCION protects ASes by providing \emph{path authorization}, \emph{valley freedom}, and \emph{loop freedom}. These security properties
rule out malicious end hosts sending packets along unauthorized, uneconomical, or impractical paths. 
We use a strong attacker model that includes end hosts colluding with \emph{compromised ASes}, whose secret MAC keys are known to the attackers. 

On the code level, we verify that the code is \emph{well-behaved}, in that it has no run-time errors, satisfies memory safety and race freedom, and processes packets in finite time. We also verify functional correctness and that the code implements the SCION protocol by verifying it against a program specification that we extract from the router model. Consequently, by Igloo's soundness guarantees, all properties that we prove for the protocol model also hold for the implementation.

\subsection{Protocol verification}
\label{ssec:protocol-verification}

The verification of SCION's data plane protocol poses several challenges: 
(1)~The security properties of the data plane must hold for arbitrary network topologies and arbitrary paths authorized by the control plane. 
(2)~Since there is one hop field per AS on a path and the path length is unbounded, the message size and number of protocol participants are unbounded.
(3)~The properties we verify are non-local properties that involve all ASes on a path, and cannot be verified by reasoning only about a single router. 
(4)~We model not only an end-host attacker, but also consider the compromise of (some) AS keys. Careful modeling is needed to exclude attacks that become unavoidable with such a strong attacker model, for instance when \emph{all} AS keys are compromised. 

Since these challenges are beyond state-of-the-art automated security protocol verifiers such as Tamarin~\cite{DBLP:conf/csfw/SchmidtMCB12,DBLP:conf/cav/MeierSCB13} and ProVerif~\cite{DBLP:conf/csfw/Blanchet01,DBLP:conf/fosad/Blanchet13}, we tackle them by formalizing the model and the properties in higher-order logic using the general-purpose Isabelle/HOL proof assistant.

We develop the SCION protocol in several steps using refinement to structure our models and proofs~\cite{dblp:journals/tcs/abadil91,DBLP:journals/iandc/LynchV95}. 
We formalize the protocol models as labeled transition systems, where each label is associated with a transition relation, called an \emph{event}. We formulate a series of such models at different abstraction levels and relate them via refinement proofs (cf.~\cref{fig:overview}): 
\begin{description}
\item[Abstract model:] contains all the essential functionality of the protocol, but no cryptography. \item[Concrete model:] introduces cryptography to protect the segments, as well as a strong attacker (see \cref{ssec:attackermodel}). \item[Decomposed model:] decomposes the monolithic concrete model into an environment model and a model for each router; also introduces separate I/O events, which correspond to I/O library calls in the implementation.
\end{description}
We express and prove the models' desired security properties as \emph{invariants}.
Since refinement preserves invariants, we can prove each security property at the most abstract model possible. This approach makes our proofs much more manageable and modular compared to proofs where one establishes the properties on the final, monolithic model.

After completing the protocol verification, we use the Igloo framework to automatically extract a program specification from the router component of the decomposed model. This \emph{I/O specification} completely describes the router's intended I/O behavior and is used as a specification against which we verify the implementation, as we discuss next.

\subsection{Code verification}
\label{subsec:overview:code-verification}

In contrast to most existing verification projects, the SCION router was developed independently from our verification effort and optimized for speed rather than verification, which poses a plethora of verification challenges. In particular, the implementation uses most of Go's language features, including some that are difficult to reason about. This includes threads, global state,
closures, and interfaces, which require advanced reasoning about concurrency, higher-order specifications, and structural subtyping. Moreover, the router implementation minimizes memory usage. The employed memory-efficient data structures and memory reclamation strategies are difficult to reason about and, to our knowledge, have not been considered in other verification projects.

We use Gobra to verify that the router implementation is well-behaved and satisfies the protocol model's I/O specification. Gobra is an SMT-based automated deductive verifier for Go that supports all Go features used by the router implementation.
Gobra verification is modular, which is crucial for verification to scale. That is, Gobra verifies each function individually, against a user-written specification. Each function specification consists of a precondition, which expresses constraints on the parameters and states in which a function may be called, and a postcondition, which describes the function results and its side effects. As usual for deductive verification tools, programmers must supply loop invariants. Specifications use separation logic~\cite{DBLP:conf/lics/Reynolds02} to express properties of the heap memory as well as the I/O performed by a function. \looseness=-1

Gobra is designed to be sound, i.e., not miss errors. 
However, it may produce spurious errors (false positives), mostly 
when the SMT solver fails to prove a valid condition. In these cases, programmers can introduce additional annotations to guide the proof search so that verification succeeds.

\subsection{Assumptions}
Our guarantees hold under the following assumptions:
\begin{description}
    \item[Environment.] We prove the security of the SCION router based on a model of the environment consisting of a network and an attacker model; we assume that these models cover the behavior of the real environment. We also assume the correctness and security of the control plane. Moreover, we verify the source code of the router implementation, but assume the correctness of the compiler, runtime system, operating system, and hardware.

    \item[Libraries.] The router implementation uses external libraries (e.g., Go's standard library), whose behaviors we specified using pre- and postconditions, but we did not verify their implementations. We did the same for a few SCION libraries that perform minor auxiliary tasks, like logging. Moreover, our proof uses a few trusted lemmas that relate the byte-level and abstract representations of packets.

    \item[Tool soundness.] We assume that the Isabelle/HOL and Gobra verification tools are sound. Their connection, given by the extraction of the I/O specification from the component model, is correct by Igloo's soundness proof, which is itself formalized in Isabelle/HOL\@. Only a small, mostly syntactic step, where we manually translate the I/O specification from Isabelle to Gobra syntax, is unverified. 

\end{description}

\section{Protocol verification}
\label{sec:protocol-verification}
We describe here our verification of the SCION data plane protocol, covering the first four levels in \cref{fig:overview}.

\subsection{Environment model}
\label{ssec:attackermodel}

\paragraph{Network and control plane model}

The state of our transition systems consists of all packets being forwarded: each packet is either in some set $\internal{A}$, the internal network of AS $A$, or in some set $\external{A,i,B,j}$, the communication channel that connects interface~$i$ of AS~$A$ to interface~$j$ of AS~$B$ via an inter-AS link. With each such link, we associate a type: \ProvCust{}, \CustProv{}, or \Core. 
We do not model intra-AS networking, as it is not managed by SCION.

To reason about segments and their combinations, our model introduces as parameter a set $auth$ of authorized segments (divided into the three types of segments: up, down and core) that represents the set of segments constructed by the control plane (cf.~\Cref{sec:scion}).

\paragraph{Attacker model} 
\looseness=-1
Our attacker can inject new packets anywhere into the network  (\internal{} and \external{}). These are then forwarded by honest router events.
In the abstract model, which does not use cryptography, the attacker can freely combine path segments to form packets, but the segments must be authorized. 
In the concrete model, we lift this limitation and introduce a full fledged Dolev-Yao attacker~\cite{dblp:journals/tit/dolevy83}, who manipulates symbolically represented messages and can eavesdrop on or inject new packets globally. 
The attacker's knowledge $K$ consists of eavesdropped messages, all authorized segments, and the secret keys of a set of compromised ASes. The attacker's forging capabilities are modeled as a closure operator $\DY(K)$, which closes the set~$K$ under the messages they can derive (e.g., by constructing MACs with known keys) and hence inject. This attacker model assumes perfect cryptography; it only considers protocol flaws relating to the \emph{use} of cryptographic primitives, not flaws in the  primitives themselves.

\subsection{Verified security properties}
\label{ssec:verified-security-properties}

We prove the following protocol-level security properties of SCION's data plane as invariants of the abstract model. 
\begin{itemize}
    \item \emph{Path authorization}:
    each segment that a packet actually traverses is contained in a segment in $auth$. \item \emph{Valley freedom}: a packet that has already traversed a \ProvCust{} link must not traverse a \CustProv{} link. \item \emph{Loop freedom}: a packet must not traverse the same link twice. \end{itemize}
We express these invariant properties in terms of a \emph{path history} associated with each packet, which records the packet's traversed path segments. This is a history variable~\cite{dblp:journals/tcs/abadil91}, used only for property specification.
As mentioned, these properties require assumptions to rule out unavoidable but trivial attacks against the compromised ASes themselves.

We leverage refinement to separate concerns. 
In the abstract model, the attacker can only choose from authorized segments, making path authorization trivial. This allows us to focus on the validity of segment combinations (valley and loop freedom). 
We then refine this model, where we show that packets still only follow authorized segments, despite the addition of a strong attacker.
Our proofs rely on auxiliary invariants: three in the abstract model, two in the concrete model, and one in the decomposed model. The refinements preserve all three main properties.

\subsection{Protocol models}
\label{ssec:protocol}

Our protocol models describe how packets are created, processed, and forwarded between the links $\external{A,i,B,j}$ and the internal networks $\internal{A}$ constituting our models' state.

We first define the abstract and concrete packets and events. We relate them in the refinement in \cref{ssec:refinement} and link our models to the code in \cref{ssec:linking-protocol-to-code-verification}. 

\subsubsection{Packet structure}  
A packet consists of up to three segments. Each segment consists of a direction flag ($\dir{}$), a \emph{segment identifier} ($\segid{}$), and a sequence of hop fields (cf.~\cref{sec:scion}). Counters are used to keep track of the current segment and hop field. 
\begin{figure}[t]
    \centering
    \includegraphics[width=\columnwidth]{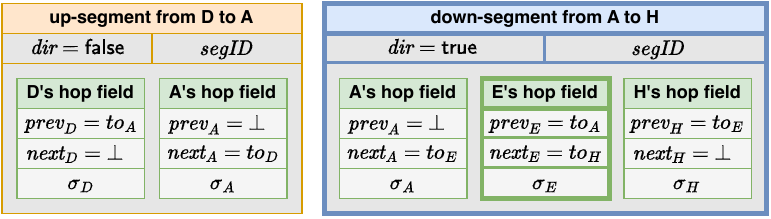}
    \caption{A two-segment packet at hop E of the path D-A-E-H in \cref{fig:overview} (payload not shown). For up-segments, where $\dir{} = \mathsf{false}$, the meaning of $\prev$ and $\nxt$ is reversed. The first hop field's $\prev{}$ and the last hop field's $\nxt{}$ are empty~($\bot$). The current segment and hop field have bold frames.
}
\label{fig:packet}
\end{figure}
\Cref{fig:packet} shows an example two-segment packet for the path D-A-E-H in \cref{fig:overview}.

Packets in the abstract and concrete model differ in their use of the segment identifier $\segid{}$ and authenticator fields $\sigma_X$.
In the abstract model, where the attacker cannot modify segments, the AS's name suffices as an authenticator and the segment identifiers are not used. In the concrete model, these fields are instantiated with cryptographic values.

\paragraph{Concrete cryptographic authenticators}  
During segment construction, each AS $X$ on a segment creates a hop field with its local forwarding information (the interfaces $\prev{}_X$ and $\nxt{}_X$) and authorizes the segment's use by embedding a cryptographic authenticator $\sigma_X$ in the hop field.
This authenticator is a message authentication code (MAC) constructed using a local MAC key $K_X$ that is shared between the border routers within AS $X$: 
\begin{equation}
\label{eq:sigma}
\sigma_X = \mathsf{MAC}_{K_X}(\prev{}_X,~\nxt{}_X,~\segid{}).\end{equation}
Here, $\segid{}$ is the (mutable) segment identifier, which the first AS on a segment initializes with a random value $RND$. Each subsequent AS $X$, after computing its $\sigma_X$, updates
\begin{equation}
\label{eq:segid-update}
\segid{} := \sigma_X \xor \segid{},
\end{equation}
where $\xor$ is exclusive-or (XOR). 
Each hop field's MAC thus protects the hop's $\prev{}$ and $\nxt{}$ fields and also, via nested MACs, all preceding hop fields. 
Subsequent hop fields are not explicitly authorized by AS $X$~\cite{KlenzeSprengerBasin-JCS-2022}.

During forwarding, AS $X$ reconstructs $\sigma_X$ using~\cref{eq:sigma}, checks that it matches the hop field's authenticator, and forwards the packet with $\segid{}$ updated as in~\cref{eq:segid-update}.

\subsubsection{Events}
\label{subsec:events}
Our models each have a \emph{router event}, which forwards packets, and an \emph{attacker event}, which injects attacker-fabricated packets into the network.

\paragraph{Router event}

We present below the event $\mathbf{\forwardevent}_a$ (resp. $\mathbf{\forwardevent}_c$) in the abstract (resp. concrete) model, which describes how a router in AS $A$ receives a packet $\pkt$ from an interface $i$, processes it, and sends it to the next hop on the path over interface $j$. 
The notation $\event{evt}{gd \,\eventsep\, upd}$ indicates that if the guard predicate $gd$ holds in the current state, then the event $evt$ updates the state as indicated by $upd$.

\begin{minipage}{.22\textwidth} \begin{align*}
& \forwardevent_a(A, \pkt, i, j):  \\
& \enskip \incheck(\pkt,A, i) \land {}  \\ 
& \enskip \ifsvalid(\pkt,A, i, j)\\
& \enskip \eventsepafter \add(\upd_a(\pkt), A, j) \\
& \enskip 
\end{align*}
\end{minipage}\hfill \begin{minipage}{.22\textwidth} \begin{align*}
& \forwardevent_c(A, \pkt, i, j):  \\
& \enskip \incheck(\pkt,A, i) \land {} \\
& \enskip \ifsvalid(\pkt,A, i, j) \land {}  \\ 
& \enskip \cryptovalid(\pkt,A)\\
& \enskip \eventsepafter \add(\upd_c(\pkt), A, j)
\end{align*}
\end{minipage}
\smallskip

The specific checks and updates performed by the above functions depend on multiple factors. 
For simplicity, we first describe these functions for a packet that is received from the internal network and forwarded to an inter-AS link, without a segment switch.
Let $\prev{}_A$ and $\nxt{}_A$ be the interfaces and $\sigma_A$ the authenticator of the current hop field, and $\segid{}$ be the current segment identifier. Then, in this case: 
\begin{itemize} \item $\incheck(\pkt, A, i)$ receives the packet $\pkt$ from the (internal) interface $i$ of AS $A$;

\item $\id{\ifsvalid{}}(\pkt,A, i, j)$ checks that the hop field interface $\nxt{}_A$ matches the router's local interface $j$;

\item $\id{crypto\_valid}$ checks that $\sigma_A$ satisfies 
\cref{eq:sigma}; 

\item $\add(\pkt, A, j)$ adds the message $\pkt$ to
$\external{A,j,B,k}$, where $B$ and $k$ are uniquely determined by $A$ and $j$;  \item $\id{upd}_a(\pkt)$ advances the hop field counter; the concrete $\id{upd}_c(\pkt)$ also updates \segid{} using \cref{eq:segid-update}. \end{itemize}

For the other cases, these functions follow a similar format, but perform different actions. 
For example, $\ifsvalid$ performs additional checks for externally received packets and for packets that switch to a new segment.

Overall the router event must consider: whether forwarding to the next AS happens on the same or a different router in the same AS, the direction of forwarding (which affects the checks done by $\cryptovalid$), whether a packet is leaving or entering an AS, and whether segment switching occurs or not. 
Our definitions allow for all combinations of these factors, which adds substantial complexity to our models.

\paragraph{Attacker event} 
In the abstract attacker event, the adversary can only create packets consisting of authorized segments ($\AuthSeg(\pkt)$). In the concrete event, we instead allow the attacker to inject arbitrary derivable packets. These events also cover the case of honest senders.

\begin{minipage}{.2\textwidth} 
\begin{align*}
& \event{\dispatchevent_a(A, \pkt, \outasinterfacevarname)}\\
& \quad \eventsepbefore AuthSeg(\pkt) \\
& \quad \eventsepafter \add(\pkt, A,\outasinterfacevarname) 
\end{align*}
\end{minipage}
\begin{minipage}{.2\textwidth} 
\begin{align*}
& \event{\dispatchevent_c(A, \pkt, \outasinterfacevarname)}\\
&\quad \eventsepbefore \pkt \in \DY(K) \\
&\quad \eventsepafter \add(\pkt, A,\outasinterfacevarname) 
\end{align*}
\end{minipage}
\smallskip

\subsection{Refinement of the abstract to the concrete model}
\label{ssec:refinement}

For the refinement proof, we define a packet abstraction function that we apply to each packet in the concrete state to obtain the corresponding abstract state. 
A crucial observation is that although the attacker can embed arbitrary hop fields, forwarding will proceed only when the hop fields have interfaces and authenticators that pass the routers' validation checks. Hence, the packet abstraction function maps the \emph{cryptographically valid prefix} of a concrete packet's path to its abstract counterpart, where each (valid) authenticator is replaced by the name of the AS that created it. We then show that (i) each concrete event's guard strengthens the corresponding abstract guard and (ii) the respective updates again lead to states related by packet abstraction. 
Point~(ii) is easy for both events, since the abstract events use the abstraction of the concrete packet. We therefore focus on point~(i), which is also straightforward for the router event, since we only \emph{add} checks in the concrete guard.  

For point (i) of the attacker event, we prove that the concrete attacker can only derive packets whose abstract counterpart is authorized. 
To this end, we show that valid derivable hop fields belong to an authorized segment. 
We first define a function to extract an abstract path from a single hop field's MAC, following its nested structure. This function is merely a construct used in the proof, not in the protocol itself (which would make it unimplementable, as actual MACs lack internal structure).
We then show that such an extracted path segment is authorized. 
Since the hop field's MAC is nested over the authorized segment, a packet's path, as long as it is valid, must correspond to this segment. 
Hence, forwarding occurs only along authorized segments. This proof involves reasoning about XOR, which is notoriously difficult~\cite{ABADI20062,ESCOBAR2012898}.

\subsection{Linking protocol verification to code verification}
\label{ssec:linking-protocol-to-code-verification}
To transition from protocol to code verification, we decompose the concrete protocol model into a model for each router and an environment model. 
We show that the composition of these models refines the original concrete model. In this refinement, we introduce buffering to separate the router's I/O operations from its internal message processing, in order to subsequently map these operations to the implementation's I/O library calls. 
Concretely, the decomposed model has events $\mathbf{\recvevent{}}_d$,  $\mathbf{\forwardevent{}}_d$, and $\mathbf{\sendevent{}}_d$, whose combination corresponds to a single $\mathbf{\forwardevent{}_c}$ event. $\mathbf{\forwardevent{}}_d$ performs checks and updates the packet, but instead of directly performing I/O, it reads from an input buffer and writes to an output buffer. I/O is performed by the other two events.

Recall that we use the Igloo methodology to link our refinement proofs in Isabelle/HOL with the Gobra code verification tool. We have extended Igloo's tooling to automatically generate the I/O specification from the routers' models within Isabelle/HOL, along with a correctness proof showing its trace equivalence with the router model.
Previously, this required a manual translation and correctness proof in Isabelle/HOL.
Finally, we express this I/O specification in Gobra's specification language, which requires a straightforward manual translation. In the next section, we show that the router conforms to this specification. 
Igloo's soundness result then guarantees that the security properties proved for the protocol also hold for the executing system.

\section{Code verification}
\label{sec:code-verification}

\looseness=-1
In this section, we explain how we verified the performance-optimized router implementation running in actual SCION deployments to be well-behaved and functionally correct, to process each incoming packet in finite time, and to correctly implement the SCION protocol.  

We verify the intended properties by annotating all functions in the router implementation with pre- and postconditions and using Gobra to check that the function implementations satisfy their specifications for all possible executions. Annotations are expressed in special comments that are recognized by the verifier, but ignored by other tools like the compiler. Verification in Gobra is modular, which allows us to split the overall verification effort into small tasks, which can be tackled in parallel. We structure  verification further by proving the intended properties in four steps:

\begin{enumerate}
\item \textit{Safety:} the implementation is well-behaved, that is, it neither crashes nor causes data races.

\item \textit{Functional correctness:} the functions compute the intended results, for instance, they perform the required checks on packet headers.

\item \textit{Finite processing:} the code for processing a SCION packet terminates.

\item \textit{I/O behavior:} the router sends SCION packets only as prescribed by the protocol model. \end{enumerate}

These steps are performed by progressively strengthening the function specifications,
which happens in a tight feedback loop with the verifier. Gobra is integrated into the IDE and verifies code incrementally during its development. Verification errors are reported to the programmer in terms of the original Go code and its annotations; the programmer never gets exposed to the underlying verification logic. 
In addition to the IDE, we have integrated verification into a continuous integration action on GitHub, verifying on every push that the entire implementation satisfies its specification.

\begin{figure}[t]
\begin{gobra}[numbers=left,numbersep=5pt]
//@$\;$decreases$\label{line:code-example:termination}$
//@$\;$req PktMem($\packetvargobra$) && DataPlaneMem(d)$\label{line:code-example:memory-pre}$
//@$\;$req $\mathit{ifs\_valid}$($\packetabstractiongobra$($\packetvargobra$),d,$\inasinterfacevargobra{}$,$\outasinterfacevargobra{}$)$\;$&&$\;$$\mathit{crypto\_valid}$($\packetabstractiongobra$($\packetvargobra$),d)$\label{line:code-example:function-pre}$
//@$\;$req Token(l$_0$) && IOSpec(l$_0$, s$_0$)$\label{line:code-example:io-pre}$
//@$\;$req $\packetvargobra$ $\in$ s$_0$.inputBuffer[$\inasinterfacevargobra{}$]$\label{line:code-example:model-pre}$
//@$\;$ens PktMem($\packetvargobra$) && DataPlaneMem(d)$\label{line:code-example:memory-post}$
//@$\;$ens $\packetabstractiongobra$($\packetvargobra$)$\;$==$\;$$\mathit{update}_c$(old($\packetabstractiongobra$($\packetvargobra$)))$\label{line:code-example:function-post}$
//@$\;$ens Token(l$_1$) && IOSpec(l$_1$, s$_1$)$\label{line:code-example:io-post}$
//@$\;$ens $\packetvargobra$ $\in$ s$_1$.outputBuffer[$\outasinterfacevargobra{}$]$\label{line:code-example:model-post}$
func$\;$$\forwardmethodgobra{}$($\packetvargobra$,d,$\inasinterfacevargobra{}$,$\outasinterfacevargobra{}$$\;$/*@,l$_0$,s$_0$$\;$@*/) /*@$\;$(l$_1$,s$_1$)$\;$@*/$\label{line:code-example:sig}$
\end{gobra}
\caption{The specification of the function \forwardmethod{}, consisting of
memory safety (Lines~\ref{line:code-example:memory-pre} and \ref{line:code-example:memory-post}), functional properties (Lines~\ref{line:code-example:function-pre} and \ref{line:code-example:function-post}), termination (Line~\ref{line:code-example:termination}), the I/O spec (Lines~\ref{line:code-example:io-pre} and \ref{line:code-example:io-post}), and properties about the I/O abstract state (Lines~\ref{line:code-example:model-pre} and \ref{line:code-example:model-post}). \code|req| and \code|ens| declare pre- and post.conditions, resp.
}
\label{fig:code-example}
\vspace{-0.5em}
\end{figure}

In the rest of this section, we provide more details on the four verification steps outlined above and illustrate them on the function \forwardmethod{} in \cref{fig:code-example}, which is a simplified version of function \processmethod{} from the router implementation. 
The function takes as arguments a packet \packetvar{} that is being processed, a structure \code|d| representing the router's configuration, as well as the ids of the input and output buffers, \code|i| and \code|j|.  It is called after the interface and cryptographic checks have been performed. The function implementation, which we omit for brevity, performs the update of the $\mathbf{\forwardevent}_d$ event of the decomposed model (see \cref{ssec:linking-protocol-to-code-verification}). We explain the function's specification below. 

\subsection{Safety}
\label{subsec:code-verification:safety}

Gobra checks that programs do not crash (for instance, due to null-pointer dereferencing or out-of-bounds accesses) and do not exhibit data races. For this purpose, it uses implicit dynamic frames~\cite{DBLP:conf/ecoop/SmansJP09}, a variant of separation logic~\cite{DBLP:conf/lics/Reynolds02, DBLP:conf/csl/OHearnRY01}. 

\paragraph{Permissions}
Separation logic expresses ownership of memory locations by associating a \emph{permission} with each location, which is created during allocation and can be transferred between different function executions. Gobra checks that a function may access a location only if it holds the corresponding permission; otherwise verification fails. 

Permissions are a powerful reasoning principle that Gobra uses for several purposes. Since there is only one permission for each location, it is not possible for two threads to access a location simultaneously, which rules out data races. Gobra's \emph{fractional permissions} \cite{Boyland03} allow concurrent read accesses while still ensuring exclusive writes. Moreover, allocating an array creates one permission per array slot. Any attempt to access an array (or slice) out of bounds is detected by Gobra since the permission for the (non-existent) slot is not available. Finally, permissions allow one to reason about side effects modularly. As long as a function holds on to the permission for a location, no other function can possibly modify this location. This allows Gobra to preserve properties of locations across calls and thread interleavings.

A function's pre- and postcondition  express which permissions the function expects from its caller and which are transferred back to the caller upon termination, as illustrated by Lines~\ref{line:code-example:memory-pre} and \ref{line:code-example:memory-post} in Figure~\ref{fig:code-example}. Predicates such as \code|PktMem| group together the permissions of entire data structures.

In Go, threads are typically synchronized using channels and locks. To reason about thread interactions, Gobra allows both channels and locks to be associated with an invariant. A channel invariant expresses properties of the messages sent over the channel and may also include permissions, such that a message may transfer permission to a location from the sender to the receiver. Locks are handled analogously.

\paragraph{Example challenges}

The router implementation is optimized for performance, which complicates verification, as we illustrate next on two examples.

First, packets have two data representations, as a raw bytestring and as a struct, which contains (1)~fields that store unmarshaled data from the bytestring and (2)~views of portions of the raw bytestring (so-called slices in Go). For (1), the implementation maintains invariants that relate the marshaled and unmarshaled data; every update of a packet must be verified to maintain these invariants. For (2), the raw bytestring and the slices stored in the struct share the same memory locations. This sharing complicates permission reasoning because the (exclusive) permissions to the shared memory are needed when updating either of the two representations. 
We solve this challenge using an idea similar to Rust's notion of borrowing~\cite{klabnik_nichols_2023}: We define one representation as the default. Before using the other representation, we perform a verification-only (\emph{ghost}) operation to borrow the necessary permissions from the default representation, and return them afterward.
At these points, we also specify how both representations are related.

Second, to avoid memory allocation and reduce the time spent on garbage collection, the router maintains a pool of allocated structs, for instance, to represent SCION paths. Whenever a path is created, a struct is retrieved from the pool, and returned afterward. Verification must reflect the corresponding permission transfers and also ensure that the structs are properly initialized before they are re-used.

\subsection{Functional correctness}
\label{subsec:code-verification:functional}

Functional properties describe the result and the state updates performed by a function. Even if the \emph{implementation} of a function operates on low-level data structures such as bytestrings, its \emph{specification} needs to express the functional properties in terms of the logical values the low-level data structures represent. This \emph{data abstraction} is essential to making specifications human-readable and maintainable.

The implicit dynamic frames logic used by Gobra allows programmers to express data abstraction via mathematical functions, written as side-effect free Go functions, that map a concrete data structure to its logical representation. 
In our example, we use the pure function \packetabstraction{} to abstract the concrete representation of a packet \packetvar{} to its logical value, that is, a message in the protocol model. This function is used in the precondition of \forwardmethod{} (Line~\ref{line:code-example:function-pre} in Figure~\ref{fig:code-example}) to express that the packet header has been validated. The postcondition at Line~\ref{line:code-example:function-post} states that \forwardmethod{} modifies the packet as prescribed by the $\mathit{update}_c$ operation of the protocol model (see \cref{subsec:events}); the \code|old| modifier allows the specification to refer to the initial value of the packet.

\subsection{Finite processing}
\label{subsec:code-verification:termination}

To prove that each packet is processed in finite time, we verify that each function involved in packet processing terminates, even though the router as a whole may run forever. To prove termination, we add a \emph{termination measure} (or \emph{ranking function}) to each loop. Gobra then checks that (1)~the termination measure strictly decreases in every loop iteration, (2)~the termination measure is always non-negative, and (3)~the loop body contains only calls and nested loops that terminate. Together, these three properties imply that the loop terminates. Functions are handled analogously; their termination measures must decrease on every 
recursive call, including indirect recursive calls. 
The function \forwardmethod{} terminates trivially since it has neither recursive calls nor loops, and calls terminating functions only. Hence, Line~\ref{line:code-example:termination} of Figure~\ref{fig:code-example} specifies the empty termination measure.

\subsection{I/O behavior}
\label{subsec:code-verification:iospec}

\begin{figure}[t]
\begin{gobra}[numbers=left,numbersep=5pt]
pred IOSpec(l,s) { 
  (forall $\asinterfacevargobra$,$\;$$\packetvargobra$ :: $\packetvargobra$ in s.outputBuffer[$\asinterfacevargobra$] ==>$\label{line:code-example:io-spec:guard}$ 
    Send(l, $\asinterfacevargobra$, $\packetvargobra$, ?l$_2$) &&$\label{line:code-example:io-spec:send-perm}$ 
    IOSpec(l$_2$, s[outputBuffer[$\asinterfacevargobra$] -= $\packetvargobra$])$\label{line:code-example:io-spec:io-post}$) &&
  ...  // other I/O permissions
}
\end{gobra}
\caption{The core predicate of the I/O specification generated from our protocol model. The predicate includes permissions for all I/O operationsprescribed by the protocol model (we show only the \code|Send| permission here). The recursive predicate application shows how the position in the protocol (\code|l|) and the model state (\code|s|) are updated when an I/O operation is performed. Variable \code|?l$_2$| is existentially quantified.}
\label{fig:code-example:io-spec}
\vspace{-0.5em}
\end{figure}

To verify I/O behavior, we follow Penninckx et al.~\cite{dblp:conf/esop/penninckx0p15} and associate a separation-logic permission with each I/O operation that may be performed by the implementation.
We annotate the relevant I/O operations of Go's I/O library to require and consume the corresponding permission. Consequently, a caller may perform the I/O operation only if it holds the permission; otherwise verification fails. 

\paragraph{I/O specification}

The I/O specification obtained from the protocol model (see \Cref{ssec:linking-protocol-to-code-verification}) is a separation logic formula that provides the I/O permissions for the entire router, that is, expresses which I/O operations the router may perform, their arguments, and in which order they may occur. Figure~\ref{fig:code-example:io-spec} shows the definition of the predicate \code|IOSpec|, the core of the I/O specification. The predicate's parameter \code|s| represents the state of the router model; its relation  to the router's concrete data structures is expressed via abstraction functions (recall \cref{subsec:code-verification:functional}). 
\code|IOSpec|'s other parameter, \code|l|, is a \emph{protocol location} that indicates the execution's current position in the protocol; it is used to specify valid sequences of I/O operations. This position is advanced whenever the router performs a protocol step, such as sending a packet.

An individual I/O permission is parameterized with the protocol locations before and after the operation, as well as the parameters and results of the operation. For instance, Line~\ref{line:code-example:io-spec:send-perm} provides the permission to send packet \packetvar{} over the \asinterface{} \asinterfacevar{}, provided that the packet is in the corresponding output buffer (Line~\ref{line:code-example:io-spec:guard}) and the protocol is at location \code|l|. Performing this operation advances the protocol to location \code|l$_2$|. The recursive instance of \code|IOSpec| reflects the new protocol location as well as the updated router state, where \packetvar{} is removed from the output buffer. The condition under which a \code|Send| operation is permitted and the modification of the router state are extracted automatically from the guard and update, respectively, of the corresponding event in the protocol model. 
The recursive predicate instance at Line~\ref{line:code-example:io-spec:io-post} lets the router perform the next I/O operation.

The router's full I/O specification, which is extracted from the protocol model, is \code|Token(l) && IOSpec(l, s$_0$)|, where \code|s$_0$| is the initial abstract state and \code|Token(l)| indicates the initial protocol location.  This specification, which is a precondition of the router's \code|main| function,  describes the permitted I/O operations of the entire router.

\begin{figure}[t]
\begin{gobra}[numbers=left,numbersep=5pt]
//@ req SktMem(conn) && PktMem($\packetvargobra$)$\label{line:code-example:send-spec:memory-pre}$
//@ req Token(l$_1$) && Send(l$_1$,conn,$\packetvargobra$,?l$_2$)$\label{line:code-example:send-spec:io-pre}$
//@ ens SktMem(conn) && PktMem($\packetvargobra$)$\label{line:code-example:send-spec:memory-post}$
//@ ens Token(l$_2$)$\label{line:code-example:send-spec:io-post}$
func (conn) Write($\packetvargobra$$\;$/*@,$\;$l$_1$$\;$@*/)
\end{gobra}
\caption{A simplified specification of the \code|Write| function of Go's I/O library package \code|net|. The function sends packet \packetvar{} over the network. The argument \code|conn| is the network socket representing the \asinterface{} that the packet is sent over. Lines~\ref{line:code-example:send-spec:memory-pre} and \ref{line:code-example:send-spec:memory-post} specify permissions for the packet and socket.
Lines~\ref{line:code-example:send-spec:io-pre} and \ref{line:code-example:send-spec:io-post} specify I/O permissions.}
\label{fig:code-example:send-spec}
\end{figure}

\paragraph{Verifying I/O properties}

Figure~\ref{fig:code-example:send-spec} shows a specification of the \code|Write| function from Go's I/O library, which forwards messages to the network. In addition to the memory permissions specified at Line~\ref{line:code-example:send-spec:memory-pre}, calling \code|Write| also requires the current protocol location to be \code|l$_1$| and a \code|Send|-permission to send the packet \packetvar{} over the \asinterface{}'s network socket \code|conn| (Line~\ref{line:code-example:send-spec:io-pre}). 
To verify the following call to \code|Write|
\begin{gobra}[numbers=none]
/*@$\;$l$_1$,s$_1$$\;$:=$\;$@*/$\forwardmethodgobra{}$($\packetvargobra{}$,t,conn$_\mathsf{in}$,conn$_\mathsf{out}$$\;$/*@,l,s$\;$@*/)$\label{line:code-example:send:call}$
conn$_\mathsf{out}$.Write($\packetvargobra{}$$\;$/*@,l$_1$$\;$@*/)$\label{line:code-example:send:send}$
\end{gobra}
we use the postconditions at Lines~\ref{line:code-example:io-post} and~\ref{line:code-example:model-post} of \forwardmethod{} (\cref{fig:code-example}), together with the definition of \code|IOSpec|, to obtain the I/O permission required by the \code|Write| operation.

An attempt to call \code|Write| \emph{without} calling \forwardmethod{} first would fail for two reasons: First, we could not establish \code|Token(l$_1$)|, i.e., the execution would be at the wrong protocol location. 
Second, we could not obtain the I/O permission from \code|IOSpec| because \code|$\packetvargobra$ in s.outputBuffer[$\asinterfacevargobra$]| would not hold, i.e., the router would not be in the expected state. This illustrates that verification enforces that both the sequence of I/O operations and the evolution of the router state precisely follow the protocol model.

\section{Main results}
\label{sec:main-results}

Our project achieved its goal: we successfully verified the SCION router all the way from high-level protocol models down to performance-optimized production code. In this section, we give an overview of our main results and the effort it took to produce them. A qualitative discussion and lessons learned are presented in the next section.

\subsection{Artifacts}
\label{ssec:artifacts}

\paragraph{Protocol}
We produced several formal protocol models of packet forwarding in SCION, together with their execution environment and attacker model, and used them to formally prove essential security properties. 
This Isabelle/HOL development consists of 16,100 LoC, and substantially extends the existing SCION formalization with 5,500 LoC~\cite{KlenzeSprengerBasin-JCS-2022}. 
Moreover, we extend Igloo with automation, adding 2,800 LoC to its formalization.

Overall, our Isabelle formalization takes five minutes on a laptop to verify.
We estimate that this development took two to three person years. 

The protocol formalization and proofs guarantee that packet forwarding in SCION as deployed today is secure, even in the presence of a strong attacker. Moreover, we expect them to be extremely useful during SCION's future evolution, for instance, to assess the impact of protocol changes on the intended security properties. The formalization will also be useful to verify SCION's control plane protocols.

\paragraph{Implementation}

We equipped the router's implementation with annotations that allow Gobra to prove memory-safety, crash- and data-race-freedom, functional correctness, finite processing, and compliance with the protocol models. We targeted the current, open-source implementation, which consists of 4,700 lines of Go code (ignoring comments and empty lines), including SCION-specific libraries, but excluding third-party libraries like the Go standard library or the library \texttt{gopacket}. We achieved our goal of verifying the deployed, performance-optimized implementation as is, except for  three small changes to work around limitations of Gobra. First, we rewrote one type declaration that uses a specific combination of Go's interfaces and Go's delegation mechanism that is not supported by Gobra. Second, we split some compound expressions, which allows us to add necessary annotations about intermediate results. Third, we rewrote some \code|range|-loops into regular \code|for|-loops, which in some cases simplifies permission-based reasoning.

Altogether, we fully verified 332 functions across 12 packages. Of these, 12 functions rely on simple, unproved lemmas relating the representation of packets as bytestrings and their logical counterpart. An additional 12 functions are verified except for a few paths (related to error handling) due to two implementation errors that we detected and reported, but that were fixed too late for us to verify the fix.  Another three functions are verified partially because of performance problems of Gobra.
We are currently closing this small gap.

In total, we added 13,400 lines of specifications and annotations, including 900 for the I/O specification and the definitions it depends on. We wrote another 2,400 lines of trusted specifications for the Go standard library and third-party libraries. The overhead of 2.8 lines of annotation per line of code is typical for SMT-based deductive verification and would be substantially higher for verification using an interactive theorem prover. Annotating and verifying the code took roughly 2.5 person years;

running Gobra on the implementation takes three hours on
a commodity laptop.

Our specifications express essential safety and security properties of the router implementation. As such, we expect them to greatly facilitate the evolution of the code base because Gobra can check automatically whether any changes to the implementation still satisfy the existing specification.

\subsection{Improvements of protocol and implementation}
\label{ssec:improvements}

Our work discovered numerous bugs and led to improvements of both the protocol and the implementation.
We provide additional detail in \cref{sec:reported-bugs}.

\paragraph{Discovered protocol attacks}
We found five attacks, including a critical attack that allowed a malicious sender to fabricate arbitrary forwarding paths, thus violating all three security properties. All vulnerabilities were resolved, except a minor one currently under discussion. 

\paragraph{Strengthened properties} 
\looseness=-1
Our formalization led us to discover an additional check that routers could perform, which allowed us to prove a stronger loop freedom property. 
In the original protocol, we could prove that a loop can exist only if there is \emph{some} compromised on-loop AS. With our improvement, loops can exist only if \emph{all} on-loop ASes are compromised. The developers implemented our recommendation.

\paragraph{Discovered implementation bugs.} 
\looseness=-1
We reported thirteen previously unknown issues in the router implementation, all of which have been confirmed by the SCION developers. So far, seven of them were fixed and three have pending fixes under discussion. The remaining three issues indicate latent bugs that do not currently affect the forwarding logic, but might in the future as the code evolves. The detected bugs affect the safety, functional correctness, and the I/O behavior of the router. While verifying safety, we found missing bounds checks, a \code{nil}-pointer dereference,

and data races. Verification of functional correctness uncovered missing checks after decoding SCION paths, incorrect error handling, and the incorrect accumulation of metrics. Finally, verifying the I/O behavior detected a critical bug where the router did not reject packets simultaneously originating from and destined to the internal network. This behavior is explicitly ruled out by the SCION protocol and our models, but was allowed by the implementation. These findings demonstrate that all major steps of the code verification were effective in uncovering bugs. Remarkably, all of these bugs escaped the extensive code reviews, testing, and fuzzing that are continuously performed on the code base, in parallel with formal verification.

\section{Lessons learned}
\label{sec:lessons}

In this section, we reflect on our project and draw lessons learned for future large-scale verification efforts.

\subsection{Overall methodology} 
\label{ssec:lessons-methodology}

One of the most fundamental design decisions of our project was to follow the Igloo methodology~\cite{10.1145/3428220}, which separates the protocol verification from the code verification and soundly integrates the two parts to provide overall guarantees. This methodology worked extremely well in our project. 
First, it allowed us to work on the protocol and code verification in parallel, which greatly reduced the overall time required for the project. Igloo links the most-refined protocol model to the implementation via the I/O specification. The development of all other, more-abstract protocol models as well as the verification of all other code properties (safety, functional correctness, finite processing) are independent of each other.
Second, Igloo enables using different verification techniques and tools for the protocol and code verification, respectively. This freedom allowed us to use an expressive interactive theorem prover to verify global properties of a system of unbounded size and and a separation logic based program verifier to reason about the intricacies of a performance-optimized Go implementation. Performing both verification tasks within one framework would have been infeasible.

The Igloo methodology requires that the most-refined protocol model and the implementation agree on the I/O operations performed by a system, which  are determined by the I/O library used in the implementation. Aligning our models to the code required modeling the protocol at a high level of detail, which was time-consuming, but also provided us with a more realistic model and led us to prove additional protocol properties. The extraction of the I/O specification was effortless, due to the automation that we implemented. 
Overall, the effort of linking the protocol to the code was small compared to verifying the protocol model and the code.

\subsection{Moving targets}

Our verification effort started in 2016 when SCION was a research project. Since then, SCION has matured and been further developed by multiple stakeholders, including academic research groups from different institutions, a non-profit association~\cite{scionassociation}, and a company~\cite{Anapaya}. 
During this time, changes were made to the protocol and the implementation.
Verifying a protocol and implementation that were under continual revision was one of the main challenges we faced.

\paragraph{Protocol changes}

We were faced with multiple proposed protocol changes that would affect our model~\cite{255270}; some, including the XOR-based authorization scheme, were also implemented~\cite{scionbookv2}.
Our initial formalization would have required laborious modifications to all models in the refinement sequence for each proposed protocol change, in order to evaluate their effects on the security properties.

To adapt to (proposed) changes quickly, we instead 
build on~\cite{KlenzeSprengerBasin-JCS-2022,IsaNet-AFP-2022}, which provide a 
formalization that is generic and covers an entire class of data plane protocols. The formalism defines several parameters (such as a cryptographic check function) and conditions that are sufficient for proving security in the parameterized refinement development. The proof effort for each instance is minimal and only involves defining the parameters and proving the conditions.
While we have left this out from this paper, our models make critical use of abstraction and parameterization; hence they could be easily adapted if the protocol were to change again.
\looseness=-1

\paragraph{Implementation changes} 
Like any code base, the SCION implementation is constantly changing as developers add features, optimize performance, and---until verification is complete---fix bugs. We initially annotated a clone of the SCION repository, but the two versions quickly fell out of sync, which made it hard to communicate with the developers (for instance, when reporting bugs) and diminished the value of the annotated implementation (since it was outdated). We then switched to a different mode, where we ported changes in the SCION repository into our code base every week, such that the annotated implementation stayed in sync with the current SCION code base. We fixed the version of the code base prior to the submission in order to have a stable target.
A logical next step is to merge our annotations into the SCION code base; we are currently discussing this step with the various stakeholders. 

Porting changes to our code base requires annotating new parts and often adapting annotations of changed code, for instance, to adjust loop invariants. In this process, we benefit greatly from the fact that our program verification technique verifies each function independently. This modularity confines the adaptations to a local scope and avoids adapting or re-verifying the unaffected parts of the code base.

\paragraph{Early verification}

The difficulties in verifying a system under development described above could have been avoided by performing the verification only after the protocol and implementation were stable. Nevertheless, verifying the protocol and code already \emph{during} their development has substantial benefits. 
As described in \cref{ssec:improvements}, we were able to uncover protocol vulnerabilities and make suggestions that strengthen the obtained security guarantees. Because this happened when the protocol was still relatively easy to update, all of our suggestions were implemented. In contrast, updating protocols that are already widely deployed is notoriously hard.
A similar argument applies to the implementation. Detecting bugs early reduces the risk of their exploitation by attackers, causing economic and reputation damage; the latter is especially critical for a next-generation Internet architecture that is trying to achieve widespread acceptance. Moreover, active code bases are often improved and extended, so dealing with changes is inevitable.

\subsection{Modularity}

A key success factor of our project was the decomposition of the overall effort into manageable chunks, which allowed us to work in parallel, focus on one problem at a time, reduce the impact of changes, and obtain better tool performance. We used four strategies to achieve this decomposition:

\begin{enumerate}
\item \textit{Separation of protocol and implementation:} We used the Igloo methodology to verify the protocol and its implementation separately and link the two parts soundly (see \cref{ssec:lessons-methodology}).

\item \textit{Protocol refinement:} We use refinement to develop the protocol model step by step and focus on one protocol aspect at a time (see \cref{ssec:protocol-verification}).

\item \textit{Code properties:} We annotated and verified the implementation in four layers, each focusing on a different aspect of the intended behavior (see \cref{sec:code-verification}).

\item \textit{Program structure:} We use modular program verification, which verifies each function of the program independently (see \cref{subsec:overview:code-verification}). 
\end{enumerate}

Modularity was absolutely crucial to tackle a verification project of the size and complexity considered here.

\subsection{Tool performance}

The single biggest challenge for the code verification was the performance of the verification tool. Even though Gobra is a state-of-the-art verifier, verification times were often too long to make effective progress. This observation is perhaps surprising for modular verification, where each function is verified separately and, thus, the time required to verify a function depends mostly on the size of the function and the complexity of the properties to be verified. 

The SCION router implementation contains several large functions (e.g., function \code|Run| has 96 LoC and \code|prepareSCMP| has 101 LoC) for which verification was too slow. Since we did not want to restructure the code or compromise on the verified properties, we designed a new code annotation (called \emph{outline statement}) that effectively extracts a code segment into a separate function, similar to a code refactoring but \emph{only} for verification purposes.  Analogously to functions, outline statements have pre- and postconditions and are verified separately, such that they split functions into smaller logical parts without actually changing the implementation. 

Even with modular verification, declarations in the input program give rise to global declarations and (quantified) axioms in the proof obligations. For implementations of the size targeted here, these declarations and axioms may overwhelm the SMT solver as there are too many ways to instantiate quantifiers. To address this, we developed a pre-processing step for Gobra that transforms the proof obligations in two ways before they are handed to the SMT solver. First, it splits queries into smaller sub-queries. Second, it removes irrelevant assumptions based on a conservative dependency analysis. These transformations do not affect soundness.  

Finally, we modified Gobra to give users more fine-grained control to select or fine-tune the used verification algorithms on a per-function basis. As a result, it was possible, for example, to use faster but less complete algorithms for reasoning about the heap by default, but fall back to slower but more precise algorithms where needed.

All three features have been added to Gobra and can now also be applied in other verification projects.

\section{Related work}
\label{sec:related-work}

We first discuss networking-related verification efforts and then other work on verifying complex systems.

\paragraph{Networking-related verification} \looseness=-1
As mentioned in the introduction, we build on an existing protocol verification of a simplified version of SCION~\cite{KlenzeSprengerBasin-JCS-2022,IsaNet-AFP-2022}. We adopt their refinement-based approach, but extended the models to include realistic paths and proved additional security properties, namely valley freedom and loop freedom. 
Independently, Chen et al.~\cite{DBLP:journals/corr/ChenJXLZL15} model SCION in a Prolog-style declarative language for specifying networking protocols, and verify both route authenticity (control plane) and data path authenticity (data plane). The latter is weaker than path authorization, as it considers each hop separately instead of relating successively traversed hops.
Consequently, path splicing attacks are not ruled out.
Zhang et al.~\cite{DBLP:conf/ccs/ZhangJBKHP14} 
prove additional security properties provided by a SCION extension called OPT, but they do not verify the security of SCION itself. 
Moreover, the soundness of the logic they use is questionable~\cite{ccs/Cremers08}.

Arnaud et al.~\cite{DBLP:journals/iandc/ArnaudCD14,DBLP:conf/cade/ArnaudCD11} model routing protocols in a process calculus and propose two decision procedures for their analysis. 
The first one analyzes a protocol for any topology and the second one works for a given topology. They analyze two ad-hoc routing protocols from the literature.
All of these works are limited to protocol verification, whereas we also provide guarantees for the actual implementation.

As mentioned in the introduction, our work on verifying a protocol and its implementation fundamentally differs from verifying network configurations, e.g., ~\cite{DBLP:conf/aplas/Kozen14,
LiuHSSLSWCMF18,
DBLP:conf/oopsla/WeitzWTEKT16,
DBLP:conf/sigcomm/BeckettGMW17}.

\paragraph{Verified security protocols} 

The Everest project has successfully verified several security protocols, including TLS~\cite{DBLP:conf/sp/BhargavanFKPS13,DBLP:conf/sp/Delignat-Lavaud17} and QUIC~\cite{DBLP:conf/sp/Delignat-Lavaud21} at the code level.
They implement the TLS and QUIC protocols in F\#~\cite{DBLP:conf/sp/BhargavanFKPS13} and F$^{*}$~\cite{DBLP:conf/sp/Delignat-Lavaud17,DBLP:conf/sp/Delignat-Lavaud21}, respectively, and use the associated refinement type checker (F7 and F$^{*}$, respectively) to perform the cryptographic security proofs. 
As they use a cryptographic attacker model, their proofs yield stronger security guarantees than are possible in  a Dolev-Yao model. The resulting protocol implementations, written in F\# or in OCaml extracted from~F$^{*}$, are reference implementations, which were designed for verification and achieve a significantly lower performance than OpenSSL. The authors also provide more efficient implementations written in language fragments that can be compiled to C (e.g., the Low$^*$ fragment of F$^*$~\cite{protzenkozrrwbd17}), which achieve a performance similar to an optimized implementation in some cases, but require an additional proof of equivalence with the reference implementation~\cite{DBLP:conf/sp/Delignat-Lavaud17,DBLP:conf/sp/Delignat-Lavaud21}.
In contrast, we verified a pre-existing, optimized protocol implementation and soundly connect it to a protocol model via the Igloo methodology.

Both DY*~\cite{DBLP:conf/eurosp/BhargavanBDHKSW21} and work by Arquint et~al.~\cite{DBLP:conf/ccs/ArquintSM023} verify security protocols at the code level by establishing invariants over the possible protocol traces and proving that these invariants entail the desired security properties. Our refinement-based approach increases modularity, which is crucial for verifying systems of the size and complexity of SCION\@.

\paragraph{Distributed system verification}

IronFleet~\cite{dblp:conf/sosp/hawblitzelhklpr15}, Verdi~\cite{dblp:conf/pldi/wilcoxwptwea15,dblp:conf/cpp/wooswatea16}, and Velisarios~\cite{dblp:conf/esop/rahlivvv18} respectively verify the Paxos, Raft, and PBFT consensus protocols. Chapar~\cite{dblp:conf/popl/lesanibc16} is a framework to prove the causal consistency of key-value stores. In IronFleet, both models and implementation are written in Dafny~\cite{leino10} and verified using an SMT solver. The other algorithms are modeled and proven correct in Coq. These works all use code extraction (from either Dafny or Coq), which is unsuitable for the verification of existing implementations written in commonly-used programming languages.

\paragraph{Verified centralized systems}

The seL4 operating system~\cite{kleinehacdeeknstw09} was verified in Isabelle/HOL using three abstraction levels (Isabelle, Haskell and C, the latter two imported into Isabelle) that are related by refinement proofs. The seL4 design tries to largely avoid concurrency to simplify the verification.  
CertiKOS~\cite{DBLP:conf/osdi/GuSCWKSC16} is a general-purpose OS kernel using fine-grained concurrency with modules of concurrent objects written in C or assembler, for which semantics and sophisticated verification techniques are embedded in Coq. 
While these approaches benefit from a unified verification framework (Isabelle or Coq), the resulting proofs require more manual work than our SMT-based code verification. In both projects, the code and the verification were developed by the same or closely collaborating teams, whereas we reason about all intricacies of an existing code base. 

Li et~al.~\cite{DBLP:conf/sp/LiLGNH21} verify security properties of the Linux KVM hypervisor by decomposing it into a core and several untrusted services. A Coq proof establishes that the core refines its specification, even if the untrusted services behave maliciously. In contrast, our goal was to verify an existing implementation without substantial alterations, and to reduce the verification effort by using an SMT-based code verifier.

\section{Conclusion}
\label{sec:conclusion}

We formally verified the SCION next-generation Internet router from its high-level design down to its performance-optimized implementation.
We discovered both design and implementation errors  that evaded all prior reviews and testing. A key success factor was our highly modular verification approach, which allowed us to reduce the verification complexity, work in parallel on different aspects of the problem, and confine the impact of protocol and implementation changes. 
Future work includes a tighter integration of code development and verification, incorporating upcoming features of the SCION router, and verifying the control plane.

\bibliographystyle{plain}
\bibliography{bibliography}

\begin{thebibliography}{10}

\bibitem{dblp:journals/tcs/abadil91}
Mart{\'{\i}}n Abadi and Leslie Lamport.
\newblock The existence of refinement mappings.
\newblock {\em Theor. Comput. Sci.}, 82(2), 1991.

\bibitem{ABADI20062}
Martín Abadi and Véronique Cortier.
\newblock Deciding knowledge in security protocols under equational theories.
\newblock {\em Theoretical Computer Science}, 367(1):2--32, 2006.

\bibitem{Anapaya}
{Anapaya Systems}.
\newblock \url{https://www.anapaya.net}, 2023.

\bibitem{DBLP:conf/cade/ArnaudCD11}
Mathilde Arnaud, V{\'{e}}ronique Cortier, and St{\'{e}}phanie Delaune.
\newblock Deciding security for protocols with recursive tests.
\newblock In Nikolaj Bjørner and Viorica Sofronie{-}Stokkermans, editors, {\em
  Automated Deduction - {CADE-23} - 23rd International Conference on Automated
  Deduction, Wroclaw, Poland, July 31 - August 5, 2011. Proceedings}, volume
  6803 of {\em Lecture Notes in Computer Science}, pages 49--63. Springer,
  2011.

\bibitem{DBLP:journals/iandc/ArnaudCD14}
Mathilde Arnaud, V{\'{e}}ronique Cortier, and St{\'{e}}phanie Delaune.
\newblock Modeling and verifying ad hoc routing protocols.
\newblock {\em Inf. Comput.}, 238:30--67, 2014.

\bibitem{DBLP:conf/ccs/ArquintSM023}
Linard Arquint, Malte Schwerhoff, Vaibhav Mehta, and Peter M{\"{u}}ller.
\newblock A generic methodology for the modular verification of security
  protocol implementations.
\newblock In Weizhi Meng, Christian~Damsgaard Jensen, Cas Cremers, and Engin
  Kirda, editors, {\em Proceedings of the 2023 {ACM} {SIGSAC} Conference on
  Computer and Communications Security, {CCS} 2023, Copenhagen, Denmark,
  November 26-30, 2023}, pages 1377--1391. {ACM}, 2023.

\bibitem{DBLP:conf/sigcomm/BeckettGMW17}
Ryan Beckett, Aarti Gupta, Ratul Mahajan, and David Walker.
\newblock A general approach to network configuration verification.
\newblock In {\em Proceedings of the Conference of the {ACM} Special Interest
  Group on Data Communication, {SIGCOMM} 2017}, pages 155--168. {ACM}, 2017.

\bibitem{DBLP:conf/eurosp/BhargavanBDHKSW21}
Karthikeyan Bhargavan, Abhishek Bichhawat, Quoc~Huy Do, Pedram Hosseyni, Ralf
  K{\"{u}}sters, Guido Schmitz, and Tim W{\"{u}}rtele.
\newblock {DY}*: {A} modular symbolic verification framework for executable
  cryptographic protocol code.
\newblock In {\em {IEEE} European Symposium on Security and Privacy, EuroS{\&}P
  2021, Vienna, Austria, September 6-10, 2021}, pages 523--542. {IEEE}, 2021.

\bibitem{DBLP:conf/sp/BhargavanFKPS13}
Karthikeyan Bhargavan, C{\'{e}}dric Fournet, Markulf Kohlweiss, Alfredo
  Pironti, and Pierre{-}Yves Strub.
\newblock Implementing {TLS} with verified cryptographic security.
\newblock In {\em 2013 {IEEE} Symposium on Security and Privacy, {SP} 2013,
  Berkeley, CA, USA, May 19-22, 2013}, pages 445--459. {IEEE} Computer Society,
  2013.

\bibitem{DBLP:conf/csfw/Blanchet01}
Bruno Blanchet.
\newblock An efficient cryptographic protocol verifier based on {Prolog} rules.
\newblock In {\em 14th {IEEE} Computer Security Foundations Workshop {(CSFW-14}
  2001), 11-13 June 2001, Cape Breton, Nova Scotia, Canada}, pages 82--96.
  {IEEE} Computer Society, 2001.

\bibitem{DBLP:conf/fosad/Blanchet13}
Bruno Blanchet.
\newblock Automatic verification of security protocols in the symbolic model:
  The verifier {ProVerif}.
\newblock In Alessandro Aldini, Javier L{\'{o}}pez, and Fabio Martinelli,
  editors, {\em Foundations of Security Analysis and Design {VII} - {FOSAD}
  2012/2013 Tutorial Lectures}, volume 8604 of {\em Lecture Notes in Computer
  Science}, pages 54--87. Springer, 2013.

\bibitem{Boyland03}
John Boyland.
\newblock Checking interference with fractional permissions.
\newblock In Radhia Cousot, editor, {\em Static Analysis, 10th International
  Symposium, {SAS} 2003, San Diego, CA, USA, June 11-13, 2003, Proceedings},
  volume 2694 of {\em Lecture Notes in Computer Science}, pages 55--72.
  Springer, 2003.

\bibitem{DBLP:journals/corr/ChenJXLZL15}
Chen Chen, Limin Jia, Hao Xu, Cheng Luo, Wenchao Zhou, and Boon~Thau Loo.
\newblock A program logic for verifying secure routing protocols.
\newblock {\em Logical Methods in Computer Science}, 11(4), 2015.

\bibitem{scionbookv2}
Laurent Chuat, Markus Legner, David Basin, David Hausheer, Samuel Hitz, Peter
  M\"{u}ller, and Adrian Perrig.
\newblock {\em The Complete Guide to {SCION}}.
\newblock Springer, 2022.

\bibitem{ccs/Cremers08}
Cas J.~F. Cremers.
\newblock On the protocol composition logic {PCL}.
\newblock In Masayuki Abe and Virgil~D. Gligor, editors, {\em ASIACCS}, pages
  66--76. ACM, 2008.

\bibitem{DBLP:conf/sp/Delignat-Lavaud17}
Antoine Delignat{-}Lavaud, C{\'{e}}dric Fournet, Markulf Kohlweiss, Jonathan
  Protzenko, Aseem Rastogi, Nikhil Swamy, Santiago~Zanella B{\'{e}}guelin,
  Karthikeyan Bhargavan, Jianyang Pan, and Jean~Karim Zinzindohoue.
\newblock Implementing and proving the {TLS} 1.3 record layer.
\newblock In {\em 2017 {IEEE} Symposium on Security and Privacy, {SP} 2017, San
  Jose, CA, USA, May 22-26, 2017}, pages 463--482. {IEEE} Computer Society,
  2017.

\bibitem{DBLP:conf/sp/Delignat-Lavaud21}
Antoine Delignat{-}Lavaud, C{\'{e}}dric Fournet, Bryan Parno, Jonathan
  Protzenko, Tahina Ramananandro, Jay Bosamiya, Joseph Lallemand, Itsaka
  Rakotonirina, and Yi~Zhou.
\newblock A security model and fully verified implementation for the {IETF}
  {QUIC} record layer.
\newblock In {\em 42nd {IEEE} Symposium on Security and Privacy, {SP} 2021, San
  Francisco, CA, USA, 24-27 May 2021}, pages 1162--1178. {IEEE}, 2021.

\bibitem{dblp:journals/tit/dolevy83}
Danny Dolev and Andrew~Chi{-}Chih Yao.
\newblock On the security of public key protocols.
\newblock {\em {IEEE} Trans. Information Theory}, 29(2), 1983.

\bibitem{ESCOBAR2012898}
Santiago Escobar, Ralf Sasse, and José Meseguer.
\newblock Folding variant narrowing and optimal variant termination.
\newblock {\em The Journal of Logic and Algebraic Programming}, 81(7):898--928,
  2012.

\bibitem{DBLP:conf/osdi/GuSCWKSC16}
Ronghui Gu, Zhong Shao, Hao Chen, Xiongnan~(Newman) Wu, Jieung Kim, Vilhelm
  Sj{\"{o}}berg, and David Costanzo.
\newblock {CertiKOS}: An extensible architecture for building certified
  concurrent {OS} kernels.
\newblock In Kimberly Keeton and Timothy Roscoe, editors, {\em 12th {USENIX}
  Symposium on Operating Systems Design and Implementation, {OSDI} 2016,
  Savannah, GA, USA, November 2-4, 2016}, pages 653--669. {USENIX} Association,
  2016.

\bibitem{dblp:conf/sosp/hawblitzelhklpr15}
Chris Hawblitzel, Jon Howell, Manos Kapritsos, Jacob~R. Lorch, Bryan Parno,
  Michael~L. Roberts, Srinath T.~V. Setty, and Brian Zill.
\newblock Ironfleet: proving practical distributed systems correct.
\newblock In Ethan~L. Miller and Steven Hand, editors, {\em Proceedings of the
  25th Symposium on Operating Systems Principles, {SOSP} 2015, Monterey, CA,
  USA, October 4-7, 2015}. {ACM}, 2015.

\bibitem{klabnik_nichols_2023}
Steve Klabnik and Carol Nichols.
\newblock {\em The {Rust} programming language}.
\newblock No Starch Press, 2023.

\bibitem{kleinehacdeeknstw09}
Gerwin Klein, Kevin Elphinstone, Gernot Heiser, June Andronick, David Cock,
  Philip Derrin, Dhammika Elkaduwe, Kai Engelhardt, Rafal Kolanski, Michael
  Norrish, Thomas Sewell, Harvey Tuch, and Simon Winwood.
\newblock {seL4}: formal verification of an {OS} kernel.
\newblock In Jeanna~Neefe Matthews and Thomas~E. Anderson, editors, {\em
  Symposium on Operating Systems Principles ({SOSP})}. {ACM}, 2009.

\bibitem{IsaNet-AFP-2022}
Tobias Klenze and Christoph Sprenger.
\newblock {IsaNet}: Formalization of a verification framework for secure data
  plane protocols.
\newblock {\em Archive of Formal Proofs}, June 2022.
\newblock \url{https://isa-afp.org/entries/IsaNet.html}, Formal proof
  development.

\bibitem{klenzecsf}
Tobias Klenze, Christoph Sprenger, and David Basin.
\newblock Formal verification of secure forwarding protocols.
\newblock In {\em 2021 IEEE 34rd Computer Security Foundations Symposium
  (CSF)}. IEEE, 2021.

\bibitem{KlenzeSprengerBasin-JCS-2022}
Tobias Klenze, Christoph Sprenger, and David Basin.
\newblock {IsaNet}: A framework for verifying secure data plane protocols.
\newblock {\em Journal of Computer Security}, 2022.

\bibitem{DBLP:conf/aplas/Kozen14}
Dexter Kozen.
\newblock {NetKAT} -- {A} formal system for the verification of networks.
\newblock In {\em Programming Languages and Systems - 12th Asian Symposium,
  {APLAS} 2014}, pages 1--18, 2014.

\bibitem{DBLP:conf/popl/KumarMNO14}
Ramana Kumar, Magnus~O. Myreen, Michael Norrish, and Scott Owens.
\newblock {CakeML}: a verified implementation of {ML}.
\newblock In Suresh Jagannathan and Peter Sewell, editors, {\em The 41st Annual
  {ACM} {SIGPLAN-SIGACT} Symposium on Principles of Programming Languages,
  {POPL} '14, San Diego, CA, USA, January 20-21, 2014}, pages 179--192. {ACM},
  2014.

\bibitem{255270}
Markus Legner, Tobias Klenze, Marc Wyss, Christoph Sprenger, and Adrian Perrig.
\newblock {EPIC}: Every packet is checked in the data plane of a path-aware
  internet.
\newblock In {\em 29th {USENIX} Security Symposium ({USENIX Security})}, pages
  541--558. {USENIX} Association, August 2020.

\bibitem{leino10}
K.~Rustan~M. Leino.
\newblock Dafny: An automatic program verifier for functional correctness.
\newblock In Edmund~M. Clarke and Andrei Voronkov, editors, {\em Logic for
  Programming, Artificial Intelligence, and Reasoning (LPAR)}, volume 6355.
  Springer, 2010.

\bibitem{DBLP:journals/cacm/Leroy09}
Xavier Leroy.
\newblock Formal verification of a realistic compiler.
\newblock {\em Commun. {ACM}}, 52(7):107--115, 2009.

\bibitem{dblp:conf/popl/lesanibc16}
Mohsen Lesani, Christian~J. Bell, and Adam Chlipala.
\newblock Chapar: certified causally consistent distributed key-value stores.
\newblock In Rastislav Bod{\'{\i}}k and Rupak Majumdar, editors, {\em
  Proceedings of the 43rd Annual {ACM} {SIGPLAN-SIGACT} Symposium on Principles
  of Programming Languages, {POPL} 2016, St. Petersburg, FL, USA, January 20 -
  22, 2016}. {ACM}, 2016.

\bibitem{DBLP:conf/sp/LiLGNH21}
Shih{-}Wei Li, Xupeng Li, Ronghui Gu, Jason Nieh, and John~Zhuang Hui.
\newblock A secure and formally verified linux {KVM} hypervisor.
\newblock In {\em 42nd {IEEE} Symposium on Security and Privacy, {SP} 2021, San
  Francisco, CA, USA, 24-27 May 2021}, pages 1782--1799. {IEEE}, 2021.

\bibitem{LiuHSSLSWCMF18}
Jed Liu, William~T. Hallahan, Cole Schlesinger, Milad Sharif, Jeongkeun Lee,
  Robert Soul{\'{e}}, Han Wang, Calin Cascaval, Nick McKeown, and Nate Foster.
\newblock p4v: practical verification for programmable data planes.
\newblock In Sergey Gorinsky and J{\'{a}}nos Tapolcai, editors, {\em
  Proceedings of the 2018 Conference of the {ACM} Special Interest Group on
  Data Communication, {SIGCOMM} 2018, Budapest, Hungary, August 20-25, 2018},
  pages 490--503. {ACM}, 2018.

\bibitem{DBLP:journals/iandc/LynchV95}
Nancy~A. Lynch and Frits~W. Vaandrager.
\newblock Forward and backward simulations: I. untimed systems.
\newblock {\em Inf. Comput.}, 121(2), 1995.

\bibitem{DBLP:conf/cav/MeierSCB13}
Simon Meier, Benedikt Schmidt, Cas Cremers, and David~A. Basin.
\newblock The {TAMARIN} prover for the symbolic analysis of security protocols.
\newblock In Natasha Sharygina and Helmut Veith, editors, {\em Computer Aided
  Verification - 25th International Conference, {CAV} 2013, Saint Petersburg,
  Russia, July 13-19, 2013. Proceedings}, volume 8044 of {\em Lecture Notes in
  Computer Science}, pages 696--701. Springer, 2013.

\bibitem{dblp:books/sp/nipkowpw02}
Tobias Nipkow, Lawrence~C. Paulson, and Markus Wenzel.
\newblock {\em {Isabelle/HOL} - A Proof Assistant for Higher-Order Logic},
  volume 2283.
\newblock Springer, 2002.

\bibitem{DBLP:conf/csl/OHearnRY01}
Peter~W. O'Hearn, John~C. Reynolds, and Hongseok Yang.
\newblock Local reasoning about programs that alter data structures.
\newblock In {\em {CSL}}, volume 2142 of {\em Lecture Notes in Computer
  Science}, pages 1--19. Springer, 2001.

\bibitem{dblp:conf/esop/penninckx0p15}
Willem Penninckx, Bart Jacobs, and Frank Piessens.
\newblock Sound, modular and compositional verification of the input/output
  behavior of programs.
\newblock In Jan Vitek, editor, {\em Programming Languages and Systems - 24th
  European Symposium on Programming, {ESOP} 2015, Held as Part of the European
  Joint Conferences on Theory and Practice of Software, {ETAPS} 2015, London,
  UK, April 11-18, 2015. Proceedings}, volume 9032. Springer, 2015.

\bibitem{dblp:series/isc/perrigsrc17}
Adrian Perrig, Pawel Szalachowski, Raphael~M. Reischuk, and Laurent Chuat.
\newblock {\em {SCION}: A Secure {Internet} Architecture}.
\newblock Springer, 2017.

\bibitem{DBLP:conf/sp/ProtzenkoPFHPBB20}
Jonathan Protzenko, Bryan Parno, Aymeric Fromherz, Chris Hawblitzel, Marina
  Polubelova, Karthikeyan Bhargavan, Benjamin Beurdouche, Joonwon Choi, Antoine
  Delignat{-}Lavaud, C{\'{e}}dric Fournet, Natalia Kulatova, Tahina
  Ramananandro, Aseem Rastogi, Nikhil Swamy, Christoph~M. Wintersteiger, and
  Santiago~Zanella B{\'{e}}guelin.
\newblock {EverCrypt}: {A} fast, verified, cross-platform cryptographic
  provider.
\newblock In {\em 2020 {IEEE} Symposium on Security and Privacy, {SP} 2020, San
  Francisco, CA, USA, May 18-21, 2020}, pages 983--1002. {IEEE}, 2020.

\bibitem{protzenkozrrwbd17}
Jonathan Protzenko, Jean~Karim Zinzindohou{\'{e}}, Aseem Rastogi, Tahina
  Ramananandro, Peng Wang, Santiago~Zanella B{\'{e}}guelin, Antoine
  Delignat{-}Lavaud, Catalin Hritcu, Karthikeyan Bhargavan, C{\'{e}}dric
  Fournet, and Nikhil Swamy.
\newblock Verified low-level programming embedded in {F}.
\newblock {\em {PACMPL}}, 1({ICFP}), 2017.

\bibitem{dblp:conf/esop/rahlivvv18}
Vincent Rahli, Ivana Vukotic, Marcus V{\"{o}}lp, and Paulo Jorge~Esteves
  Ver{\'{\i}}ssimo.
\newblock Velisarios: Byzantine fault-tolerant protocols powered by coq.
\newblock In Amal Ahmed, editor, {\em Programming Languages and Systems - 27th
  European Symposium on Programming, {ESOP} 2018, Held as Part of the European
  Joint Conferences on Theory and Practice of Software, {ETAPS} 2018,
  Thessaloniki, Greece, April 14-20, 2018, Proceedings}, volume 10801.
  Springer, 2018.

\bibitem{DBLP:conf/lics/Reynolds02}
John~C. Reynolds.
\newblock Separation logic: {A} logic for shared mutable data structures.
\newblock In {\em {LICS}}, pages 55--74. {IEEE} Computer Society, 2002.

\bibitem{DBLP:conf/csfw/SchmidtMCB12}
Benedikt Schmidt, Simon Meier, Cas J.~F. Cremers, and David~A. Basin.
\newblock Automated analysis of {Diffie-Hellman} protocols and advanced
  security properties.
\newblock In Stephen Chong, editor, {\em 25th {IEEE} Computer Security
  Foundations Symposium, {CSF} 2012, Cambridge, MA, USA, June 25-27, 2012},
  pages 78--94. {IEEE} Computer Society, 2012.

\bibitem{scionassociation}
{SCION Association}.
\newblock \url{https://www.scion.org}, 2023.

\bibitem{SSFN-SIX}
SIX.
\newblock {Secure Swiss Finance Network}.
\newblock
  \url{https://www.six-group.com/en/products-services/banking-services/ssfn.html}.

\bibitem{DBLP:conf/ecoop/SmansJP09}
Jan Smans, Bart Jacobs, and Frank Piessens.
\newblock Implicit dynamic frames: Combining dynamic frames and separation
  logic.
\newblock In {\em {ECOOP}}, volume 5653 of {\em Lecture Notes in Computer
  Science}, pages 148--172. Springer, 2009.

\bibitem{10.1145/3428220}
Christoph Sprenger, Tobias Klenze, Marco Eilers, Felix~A. Wolf, Peter
  M{\"{u}}ller, Martin Clochard, and David~A. Basin.
\newblock Igloo: soundly linking compositional refinement and separation logic
  for distributed system verification.
\newblock {\em Proc. {ACM} Program. Lang.}, 4({OOPSLA}):152:1--152:31, 2020.

\bibitem{DBLP:conf/oopsla/WeitzWTEKT16}
Konstantin Weitz, Doug Woos, Emina Torlak, Michael~D. Ernst, Arvind
  Krishnamurthy, and Zachary Tatlock.
\newblock Scalable verification of border gateway protocol configurations with
  an {SMT} solver.
\newblock In {\em Proc. ACM Program. Lang.}, OOPSLA 2016, page 765–780, New
  York, NY, USA, 2016. Association for Computing Machinery.

\bibitem{dblp:conf/pldi/wilcoxwptwea15}
James~R. Wilcox, Doug Woos, Pavel Panchekha, Zachary Tatlock, Xi~Wang,
  Michael~D. Ernst, and Thomas~E. Anderson.
\newblock Verdi: a framework for implementing and formally verifying
  distributed systems.
\newblock In David Grove and Steve Blackburn, editors, {\em Proceedings of the
  36th {ACM} {SIGPLAN} Conference on Programming Language Design and
  Implementation, Portland, OR, USA, June 15-17, 2015}. {ACM}, 2015.

\bibitem{DBLP:conf/cav/WolfACOPM21}
Felix~A. Wolf, Linard Arquint, Martin Clochard, Wytse Oortwijn,
  Jo{\~{a}}o~Carlos Pereira, and Peter M{\"{u}}ller.
\newblock Gobra: Modular specification and verification of go programs.
\newblock In {\em {CAV} {(1)}}, volume 12759 of {\em Lecture Notes in Computer
  Science}, pages 367--379. Springer, 2021.

\bibitem{dblp:conf/cpp/wooswatea16}
Doug Woos, James~R. Wilcox, Steve Anton, Zachary Tatlock, Michael~D. Ernst, and
  Thomas~E. Anderson.
\newblock Planning for change in a formal verification of the {Raft} consensus
  protocol.
\newblock In Jeremy Avigad and Adam Chlipala, editors, {\em Proceedings of the
  5th {ACM} {SIGPLAN} Conference on Certified Programs and Proofs, Saint
  Petersburg, FL, USA, January 20-22, 2016}. {ACM}, 2016.

\bibitem{DBLP:conf/ccs/ZhangJBKHP14}
Fuyuan Zhang, Limin Jia, Cristina Basescu, Tiffany~Hyun{-}Jin Kim, Yih{-}Chun
  Hu, and Adrian Perrig.
\newblock Mechanized network origin and path authenticity proofs.
\newblock In Gail{-}Joon Ahn, Moti Yung, and Ninghui Li, editors, {\em
  Proceedings of the 2014 {ACM} {SIGSAC} Conference on Computer and
  Communications Security, Scottsdale, AZ, USA, November 3-7, 2014}, pages
  346--357. {ACM}, 2014.

\bibitem{DBLP:conf/ccs/ZinzindohoueBPB17}
Jean~Karim Zinzindohou{\'{e}}, Karthikeyan Bhargavan, Jonathan Protzenko, and
  Benjamin Beurdouche.
\newblock {HACL*}: {A} verified modern cryptographic library.
\newblock In Bhavani~M. Thuraisingham, David Evans, Tal Malkin, and Dongyan Xu,
  editors, {\em Proceedings of the 2017 {ACM} {SIGSAC} Conference on Computer
  and Communications Security, {CCS} 2017, Dallas, TX, USA, October 30 -
  November 03, 2017}, pages 1789--1806. {ACM}, 2017.

\end{thebibliography}

\appendix

\section{Reported Bugs and Improvements}
\label{sec:reported-bugs}

\subsection{Protocol Vulnerabilities}
As mentioned in the paper, we found multiple protocol vulnerabilities in the early stages of this verification project, which lead to five concrete attacks (see \cref{table:protocolissues}). Three of these attacks related to subtle edge cases in the segment switching logic, highlighting the logic's complexity and the necessity of its formal verification.
The most severe attack allows an attacker to create an arbitrary forwarding path, hence violating all three security properties stated in the paper. This attack exploits multiple vulnerabilities, in particular, missing validation checks.

\begin{table*}[h!]
\begin{tabular}{@{}p{10cm}p{2cm}p{1cm}@{}}
\toprule
\textbf{Issue description} & \textbf{Confirmed} & \textbf{Fixed} \\ \midrule
Path splicing attack: Missing checks in segment switching. &  \centering \checkmark  &   { \centering \checkmark }   \\ \midrule Loop / traffic reflection attack: Missing checks in segment switching, limit on number of segments not enforced.   &    \centering \checkmark       &   {\checkmark}    \\ \midrule Arbitrary source routing attack: Missing checks in segment switching, limit on number of segments not enforce.d &        \centering \checkmark   &   {\checkmark}    \\ \midrule Path splicing attack: Incorrect handling of \code|currentHF| counter in combination with \code|Verify-Only| flag. &     \centering \checkmark      &    {\checkmark}   \\ \midrule Unauthorized use of peering links: Ingress router on last hop does not ensure that \code|EgressIF| is empty. &     \centering \checkmark      &       { \centering --- } \\ \bottomrule \end{tabular}
\caption{List of protocol attacks. The last attack has not been resolved yet, but it is minor compared to the other attacks.}
\label{table:protocolissues}
\end{table*}

We reported all these forwarding protocol vulnerabilities to the SCION developers, who resolved all but one minor issue that is currently under discussion. 
Some attacks were resolved directly, by adding additional checks to the router code, others exploited vulnerabilities in former protocol version of SCION~\cite{dblp:series/isc/perrigsrc17} that have been resolved in SCION's re-design~\cite{scionbookv2}.

\subsection{Protocol Improvement}
Originally, SCION routers checked for valleys only when switching between segments, as it was assumed that the control plane would correctly and securely construct internally valley-free segments.

While modeling and carrying out our formal proofs, we noticed that substantially stronger valley- and loop-freedom properties can be achieved if valley checks are also added to the intra-segment forwarding logic. In particular, we prove that valley-freedom holds even if \emph{all} on-path ASes are malicious. Furthermore, this allows us to prove a stronger loop-freedom property, stating that a loop can happen only if \emph{all} ASes in the loop are malicious (as opposed to \emph{at least one} AS, as we had previously).

\subsection{Reported Implementation Bugs}
\begin{table*}[h!]
\begin{tabular}{@{}p{10cm}p{2cm}p{1cm}@{}}
\toprule
\textbf{Issue description} & \textbf{Confirmed} & \textbf{Fixed} \\ \midrule
Missing bounds check in the function \code|HostFromRaw| that reads a host address from raw bytes &  \centering \checkmark  &   { \centering \checkmark }   \\ \midrule Missing bounds check in the function \code|SerializeTo| that serializes a Raw SCION path   &    \centering \checkmark       &   {\centering ---}    \\ \midrule Missing synchronization operation in the main loop of the router &        \centering \checkmark   &   {\centering ---}    \\ \midrule Improper error handling when sending a package through the internal network &     \centering \checkmark      &    {\centering ($*$)}   \\ \midrule Latent race condition when BFD packets are processed &    \centering \checkmark       &   { \centering \checkmark }    \\ \midrule Missing checks after decoding a SCION path &     \centering \checkmark      &   { \centering ($*$) }    \\ \midrule Incorrect error reporting when reversing an EPIC path &    \centering \checkmark       &   { \centering \checkmark }    \\ \midrule \code|nil|-pointer dereference when reversing an empty SCION path &        \centering \checkmark   &   { \centering \checkmark }    \\ \midrule Incorrect computation of
metrics about processed packets due to the improper reuse of heap-allocated data structures &     \centering \checkmark      &  { \centering \checkmark }     \\ \midrule Wrong operand used for comparing the dynamic type of a variable in function \code|nextHdr| &     \centering \checkmark      &   { \centering \checkmark }    \\ \midrule Function \code|AddLinkType| does not acquire the DataPlane mutex, unlike the other setters &     \centering \checkmark      &   {\centering ---}    \\ \midrule Missing checks for the number of hop fields in a SCION path &      \centering \checkmark     &   { \centering \checkmark }     \\ \midrule Router does not reject packets simultaneously originating from, and
bound to the internal network. &     \centering \checkmark      &       
{ \centering ($*$) } \\ \bottomrule \end{tabular}
\caption{List of issues reported in the official SCION repository on GitHub. All issues were confirmed by the developers. Furthermore, seven issues have been fixed~(\checkmark), and three of them have pending fixes and are under discussion~($*$).}
\label{table:issues}
\end{table*}
In \Cref{table:issues}, we list all the thirteen issues we identified and reported to the SCION developers. All issues were confirmed, and seven of them have been fixed already. Additionally, there are proposed solutions for three of the remaining issues.
Besides these issues, we identified a performance bug in the function \code|processIntraBFD|, due to the use of a \code|continue| statement instead of a \code|break| to exit early from a loop. This report, however, is not a direct consequence of our verification efforts because we are not using Gobra to reason about the performance characteristics of the program. Finally, we proposed two improvements to the SCION developers, which were accepted: first, we suggested passing a large structure by reference, instead of by value; second, we identified a loop that could run for, at most, one iteration. In this case, we replaced the loop with straight-line code. This had the positive side-effect of simplifying our proofs, eliminating the need for loop invariants and termination measures.

\end{document}